\documentclass[usenatbib]{mnras}

\usepackage[T1]{fontenc}
\usepackage{ae,aecompl}
\pdfoutput=1

\usepackage{pdflscape}

\usepackage{graphicx}	% Including figure files
\usepackage{amsmath}	% Advanced maths commands
\usepackage{amssymb}	% Extra maths symbols

\title[Kinematics of ORLs and CELs in PNe]{The kinematical behaviour of ORLs and CELs in Galactic PNe\thanks{Based on data obtained at Las Campanas Observatory, Carnegie
Institution.}\thanks{This work has used observations collected at the European Southern Observatory, Chile, proposal number ESO 090.D-0265(A) }}

\author[M. Pe\~na et al.]{
Miriam Pe\~na,$^{1}$\thanks{E-mail: miriam@astro.unam.mx}
Francisco Ruiz-Escobedo,$^{1}$
Jackeline S. Rechy-Garc{\'\i}a$^{1}$ and
\newauthor  Jorge Garc{\'\i}a-Rojas$^{2,3}$
\\
$^{1}$Instituto de Astronom{\'\i}a, Universidad Nacional Aut\'onoma de M\'exico, Apdo. Postal 70264,  Cd. de M\'exico, 04510, M\'exico \\
$^{2}$Instituto de Astrof\'isica de Canarias (IAC), E-38200 La Laguna, Tenerife, Spain\\
$^{3}$Universidad de La Laguna, Dept. Astrof\'{\i}sica. E-38206 La Laguna, Tenerife, Spain 
}

\date{Accepted XXX. Received YYY; in original form ZZZ}

\pubyear{2017}

\begin{document}
\label{firstpage}
\pagerange{\pageref{firstpage}--\pageref{lastpage}}
\maketitle

\begin{abstract}
The kinematics of the plasma in 14 PNe is analysed by measuring the expansion velocities (V$_{exp}$) of different ions as derived from their collisionally excited lines (CELs) and optical recombination lines (ORLs). V$_{exp}$ are analysed as a function of the ionisation potential of ions that at first approximation represents the distance of the ion  to the central star. In most cases the kinematics of ORLs is incompatible with the kinematics of CELs at the same ionisation potential, specially if CELs and ORLs of the same ion are considered. In general V$_{exp}$ from ORLs is lower than V$_{exp}$ from CELs indicating that, if the gas is in ionisation equilibrium, ORLs are emitted by a gas located closer to the central star. The velocity field derived from CELs shows a gradient accelerating outwards as predicted from hydrodynamic modelling of PNe ionisation structures. The velocity field derived from ORLs is different, in many cases the velocity gradient  is flatter or nonexistent and  high and low ionised species present nearly the same V$_{exp}$. In addition, the FWHM(ORLs) is usually smaller that the FWHM(CELs). Our interpretation is that ORLs are mainly emitted  by a plasma that coexists with the plasma emitting CELs, but does not fit  the ionisation structures predicted by models.  Such a plasma should have been ejected in a different event that the plasma emitting CELs.
\end{abstract}


%%%%%%%%%%%%%%%%%%%%%%%%%%%%%%%%%%%%%%%%%%%%%%%%%%

%%%%%%%%%%%%%%%%% BODY OF PAPER %%%%%%%%%%%%%%%%%%

\section{Introduction}

%All papers should start with an Introduction section, which sets the work
%in context, cites relevant earlier studies in the field by \citet{Others2013},
%and describes the problem the authors aim to solve \citep[e.g.][]{Author2012}.
Planetary nebulae are  constituted by  ionised gas surrounding a low-intermediate mass (1 M$_\odot \leq$ M $<$ 8 M$_\odot$) hot  evolved star. The gas was part of the stellar atmosphere and was ejected by the star, at low velocity of about 10 km s$^{-1}$,  in advanced stages of evolution, while the star was passing through the phases of Giant - Asymptotic Giant Branch (AGB). Thus the nebula is expanding away from the star with a velocity that depends on the ejection processes, as well as on thermal processes, the ionisation structure, and the interaction of the nebula with the stellar winds and the interstellar medium.

The ionised gas emits in lines corresponding to the ions present in it. Being a low density plasma,  collisionally excited lines (CELs) from ions of heavy elements  (C, N, O, Ne, Ar, S, and others) are conspicuous, and optical recombination lines (ORLs) of H, He and ions of heavy elements are emitted as well (the latter ones are usually faint). Physical conditions (electron temperature and density) and ionic chemical abundances can be derived from both types of emission lines. Total abundances are obtained by considering all the ions detected in the gas  and  including the not visible ions by using ionisation correction factors (ICFs) or other procedures.

As already said, ionic abundances can be derived by using CELs or ORLs (in the following ions emitting CELs are marked with square parenthesis, e.g., [\ion{O}{iii}] $\lambda$5007, while ions emitting ORLs do not show parenthesis, e.g., \ion{O}{ii} $\lambda$4649). However such abundances do not coincide for the same ion and usually the abundances from ORLs are larger than those from CELs by factors of 2 or larger. This occurs in \ion{H}{II} regions and PNe as well,  \citep[see e.g.,][and references therein]{garcia:07}. In a few PNe such a factor can rise up to 100 or more, but the mean value is around 3 \citep{mcnabb:13}. The origin of the discrepancy, known as the Abundance Discrepancy Factor (ADF) is, so far,  an open problem in astrophysics of gaseous nebulae. It has been atributed to temperature fluctuations \citep[e.g.,][]{peimbert:67,peimbert:71},  tiny metal-rich inclusions embedded in the H-rich plasma \citep[][and references therein]{liu:06}, gas inhomogeneities or other processes  \citep[for instance Kappa-distribution of electron energy,][]{nicholls:12}. The first two proposals would imply that CELs and ORLs would be emitted in zones of different temperatures in the nebula, because CELs are highly dependent on the electron temperature, thus CELs are emitted in zones of high temperature, while ORLs are much less dependent on electron temperature and they would be  emitted mainly in low temperature zones.

Since several years ago some authors have indicated that the ORLs and CELs of the same ion seem to show different kinematical behaviour or different physical distribution in the nebula. In several objects recombination lines of O$^{+2}$ appear to show lower expansion velocities than O$^{+2}$ collisionally excited lines \citep[e.g., the case of BB\,1 presented by ][]{otsuka:10}, or  the emission of ORLs of C$^{+2}$ and O$^{+2}$  
strongly peaks towards the nebular centre, such as the cases of NGC\,6153 \citep{liu:00}, NGC\,7009 \citep{luo:03},  NGC\,6720 \citep{garnett:01},  Abell 46 and Ou5 \citep{corradi:15} and NGC 6778 \citep{jones:16}. Also Hf\,2-2, a PN with a very large ADF of 70, presents the same phenomenon \citep{liu:06}. On the other hand, other PNe do not show differences between the kinematical behaviour or physical distribution of ORLs and CELs  \citep{otsuka:09}.
There is also an interesting work by \citet{barlow:06} who, from very high resolution spectra, found  that   ORLs from O$^{+2}$ were narrower than  CELs from the same ion, in two PNe. 

More recently the situation appears more complicated because \citet{richer:13} have analysed in detail spatially and velocity-resolved spectroscopy of ORLs and CELs for the same ions in NGC\,7009, finding that the lines show discrepant kinematics and location. Kinematics of ORLs  is incompatible with the ionisation structure given by CELs. These authors suggest that there is an additional plasma component and that emission of recombination lines arises from a different volume from that giving rise to the forbidden emission from the parent ions
within this nebula. 

Also \citet{garcia:16}, from direct imaging made with tunable filters of the faint ORL emission, found that the emissions of the ORLs \ion{O}{II} $\lambda$$\lambda$4649+50 and CEL  [\ion{O}{iii}] $\lambda$5007 have different spacial distribution in NGC\,6778. These authors argue  that these differences are consistent with the presence of a H-deficient gas where ORLs would be emitted and they suggest that the origin of such a H-deficient plasma may be linked to the binarity of the central star. Besides, they found that the distribution of the emission of the [\ion{O}{iii}] $\lambda$4363 line, sensitive to the electron temperature, strongly peaks in a region more centrally located than the [\ion{O}{iii}] $\lambda$5007 line, indicating a temperature gradient in the nebula.

Considering the above, we have decided to study a sample of PNe, ionised by different types of central stars ([WC]  stars, weak emission line stars, {\it wels}, and normal stars) for which we have very good quality high-resolution spectra, in order to explore the possibility that ORLs and CELs are emitted by gas at different conditions. For this, we analyse  the kinematics of the  nebular structure given by both types of lines,  their line widths and their profiles by using very high-resolution spectra, where the lines are well resolved. Most of the spectra were obtained  with the  Magellan Inamori Kyocera Echelle spectrograph (MIKE) \citep{berstein:03}, attached to the 6.5-m Magellan telescope Clay, at Las Campanas Observatory, Chile.  These data have been already analysed to derive physical conditions and chemical abundances from ORLs and CELs and to determine the ADFs for the objects \citep[][G-R2009, G-R2012, G-R2013 respectively]{garcia:09,garcia:12,garcia:13} 

Additionally, data for the PNe with normal central stars NGC\,7009 and NGC\,3918,  have been obtained from the literature or from other observations.
 
General characteristics of the analysed nebulae are presented in Table \ref{tab:objects} where we have included the morphological classification of the objects. For the compact, young PNe the morphological classification by \citet{sahai:11} has been adopted, when it is available. For other objects, the classification has been taken from different sources of the literature. Distances and radii have been obtained from the lists by \citet{frew:16}. A very uncertain age has been estimated from the ratio radius/V$_{exp}$(\ion{H}{i}) (column 12).

\begin{landscape}
\begin{table}
\caption{Characteristics of studied PNe. \label{tab:objects}}
\begin{tabular}{lrcccccccccrc}
\hline 
\hline \\
PN G & name & Star$^a$ & ADF(O$^{+2}$) & O/H$^b$ & N/O$^b$&  n$_e$&$\phi$& D$^d$ & rad$^d$ & morph.$^e$ &age$^e$ &ref$^c$ \\ 
  &  &  &  & & &cm$^{-3}$ & $''$ & kpc& (pc)  \\
\hline
002.2$-$09.4 & Cn\,1-5 & [WO\,4]p &1.90&8.79 &0.85 & 4000 & 7.0 & 4.99& 0.079 & B,o bcr(o,i) & 3100& (1)\\
003.1+02.9& Hb\,4 & [WO\,3]& 3.70& 8.70&0.70 & 6250 & 7.2 & 2.88 & 0.060& M,c bcr(i) an &  3400  & (1) \\
004.9+04.9 & M\,1-25 & [WC\,5-6]&1.51 & 8.87&0.34& 15100& 3.2  & 5.60& 0.052& E,c & 2676& (1)\\
006.8+04.1 & M\,3-15 &[WC\,4]& 2.34& 8.81 & 0.33& 8800 & 4.5 & 5.51 & 0.058& L,c bcr(c) &3500$^f$ &(1)\\
011.9+04.2 & M\,1-32& [WO\,4]p & 2.34& 8.74 & 0.50&15000& 11.0 & 3.56 & 0.074& Bipolar  &4500$^f$& (1) \\
019.4$-$05.3 & M\,1-61& {\it wels}& 1.66 &8.67 & 0.40 & 22200 & 1.8 & 6.61 &0.029 &M,c bcr & 1540& (1)\\
037.7$-$34.5& NGC\,7009 &pn& 5.00 & 8.61 & 0.31 &4370 &28.5 & 1.26 & 0.076 & Mshell an & 3800& (3)\\
278.1$-$05.9& NGC\,2867&  [WO\,2] &1.58& 8.58& 0.32& 4150 & 14.0 & 2.23 & 0.076&  Ellip. sm& 2660 &(2)\\
285.4+01.5 & Pe\,1-1 & [WO\,4]& 1.70 & 8.90 & 0.25& 31100 & 3.0 & 5.39 & 0.039 & B,c bcr & 3495 &(1)\\
292.4+04.1 & PB\,8 & [WC/WN]&2.19 & 8.77  & 0.27 &2550 & 6.6 & 5.36 & 0.085&Ellipt. & 4800 &(2)\\
294.6+04.7& NGC\,3918 & pn & 1.80 & 8.67 & 0.28 & 6200 &18.7  &1.55 & 0.068& point sym. & 4740 &(4)\\
300.7$-$02.0 & He\,2-86 & [WC\,4]   &1.95 & 8.79 & 0.67 & 23300 &3.2 & 4.62 & 0.035& M,c bcr(o) & 3770 &(1)\\
336.2$-$06.9& PC\,14 & [WO\,4] & 1.94 & 8.77&0.24& 3550& 7.2 & 5.97 & 0.087 & Bipolar  & 3940& (1) \\
355.9$-$04.2 & M\,1-30 & {\it wels}&2.14 & 8.90 & 0.49 & 8000& 3.5 & 6.49 & 0.055&  M,c,t  ps(m),h(a)& 4600 &(1) \\
\hline
\multicolumn{12}{l}{$^a$ [WR] and {\it wels} stellar types are from \citet{acker:03}, except for PB\,8 which is from \citet{todt:10}.}\\
 \multicolumn{4}{l}{$^b$ O/H and N/O abundance ratios are derived from CELs.}\\
\multicolumn{12}{l}{$^c$ References for ADF, O/H, N/O, and n$_e$ are: (1) G-R2013, (2) G-R2009, (3) Liu et al. 1995, (4) G-R2015.}\\
\multicolumn{5}{l}{$^d$ Distance and radius are from \citet{frew:16}.} \\  
\multicolumn{12}{l}{$^e$ Morphology  from \citet{sahai:11} or from the literature. Age from radius/V$_{exp}$(\ion{H}{i}).} \\ 
\multicolumn{7}{l}{ ~~~NGC\,7009 has a very complex multishell structure, with ansae.}\\
\multicolumn{12}{l}{ $^f$ For M\,1-32 and M\,3-15, the ages are from the hydrodynamical models by \citet{ of O$^{+3}$:17}.}\\
\end{tabular}
\end{table}
\end{landscape}

\section{Description of Data}

Data reduction and line measurements were performed following IRAF\footnote{IRAF is distributed by the National Optical Astronomy Observatories, which is operated by the Association of Universities for Research in Astronomy, Inc., under contract to the National Science Foundation.} standard procedures. Line intensities from MIKE  data were already presented in G-R2009 and G-R2012, where reddening and physical conditions were determined for the objects. Line intensities were used by G-R2009 and G-R2013 to derive ionic and total abundances from ORLs and CELs and to compute the ADF for each object. The ADFs for this PN sample are moderate, in a range from 1.2 to 3.7.

\begin{figure*}
\begin{center}
\caption{Line profiles of the auroral CEL [\ion{O}{iii}] $\lambda$4363 and the ORL \ion{O}{ii} $\lambda$4649 of PNe analysed in this work (NGC\,7009 is not shown). \label{fig:profiles}}
\includegraphics[scale=0.20]{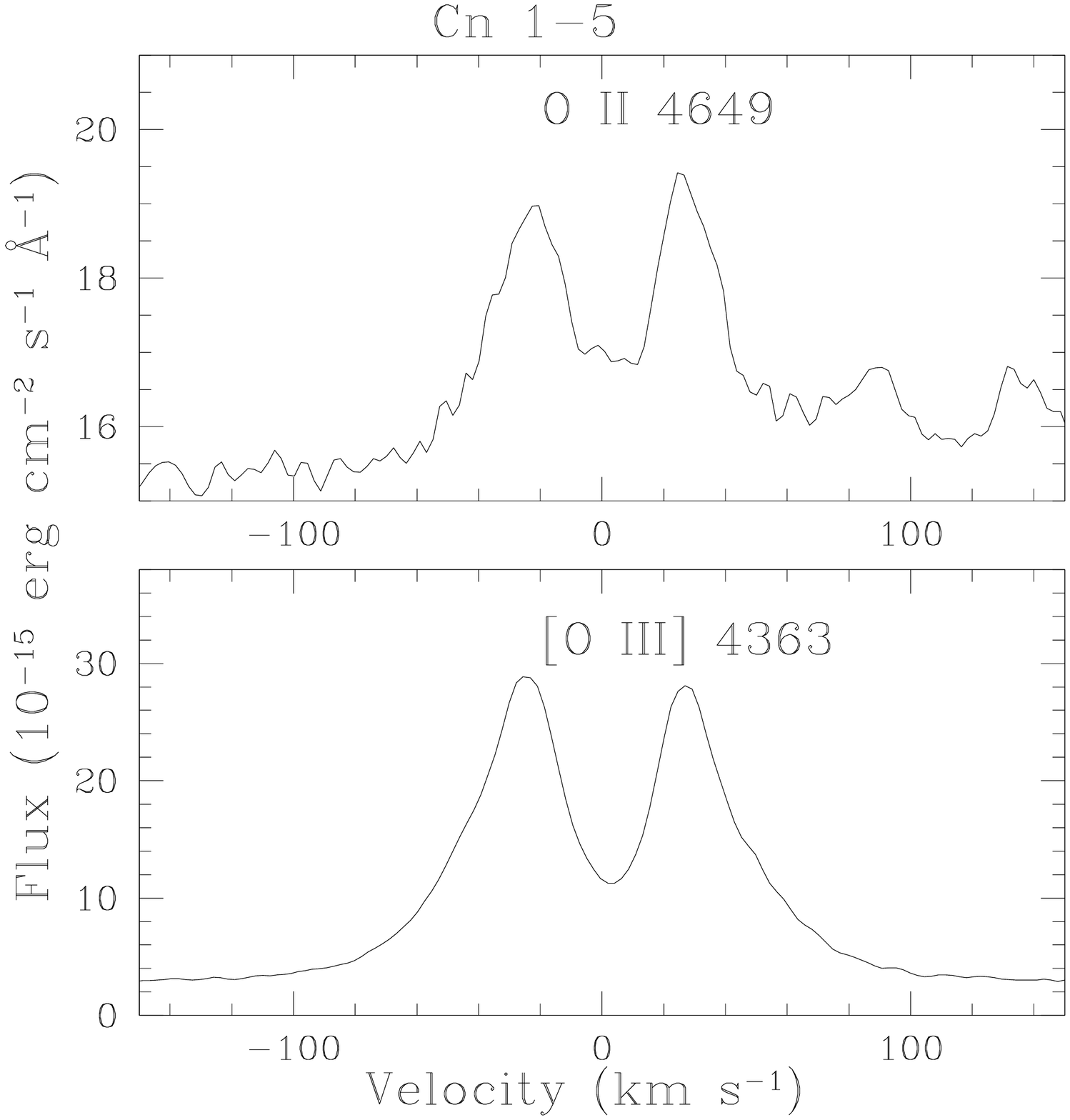}
\includegraphics[scale=0.20]{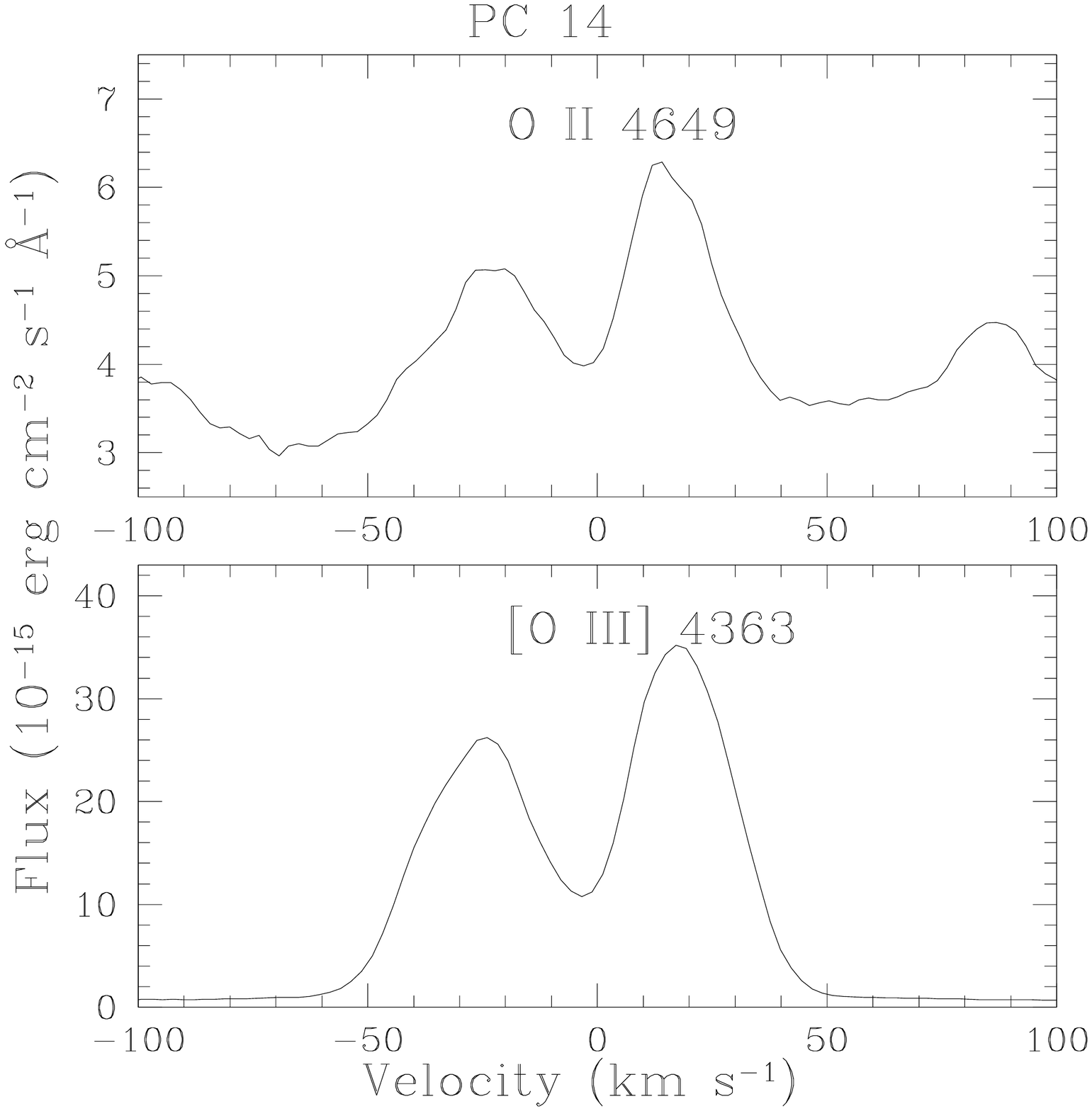}
\includegraphics[scale=0.20]{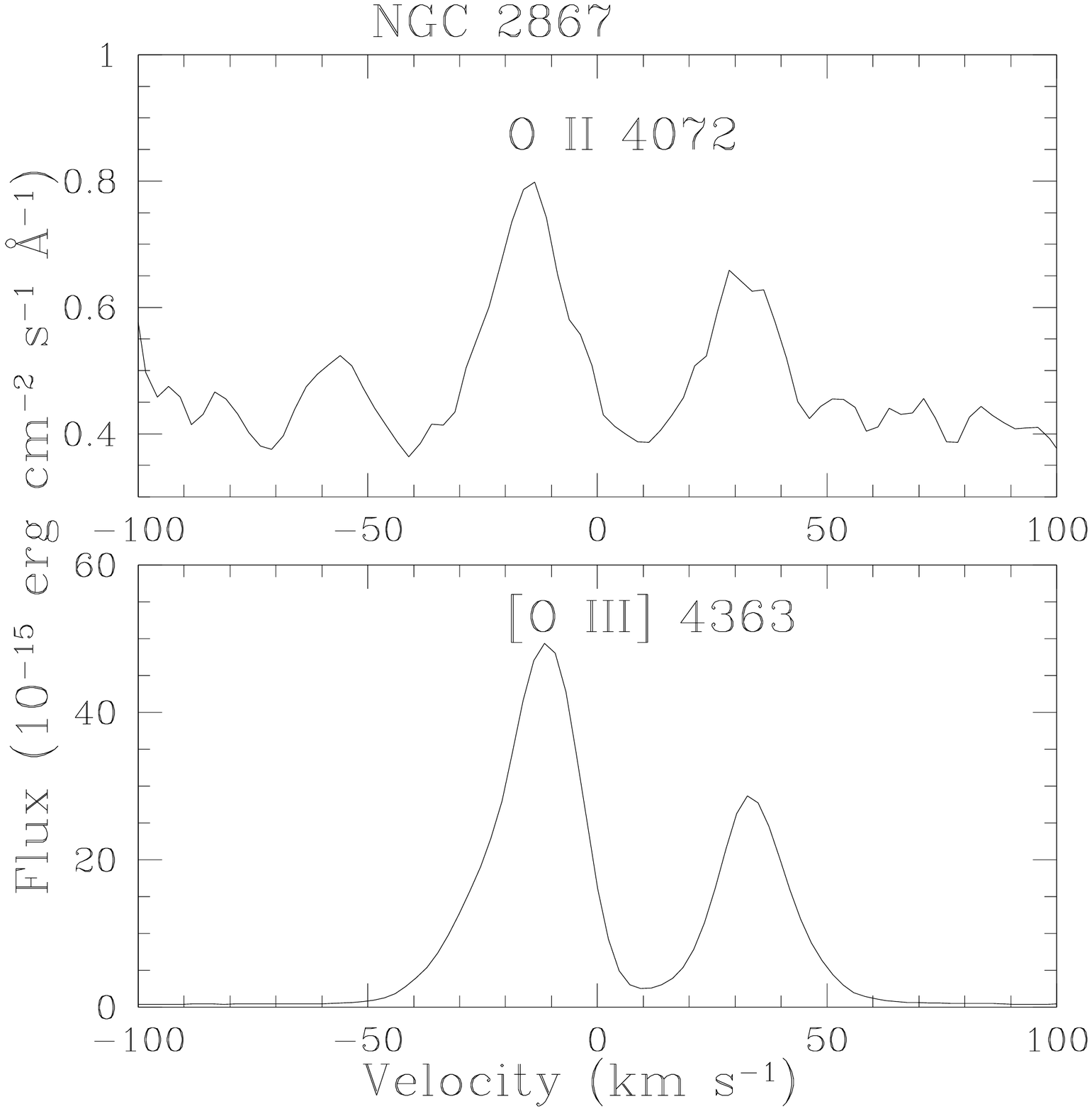}
\includegraphics[scale=0.20]{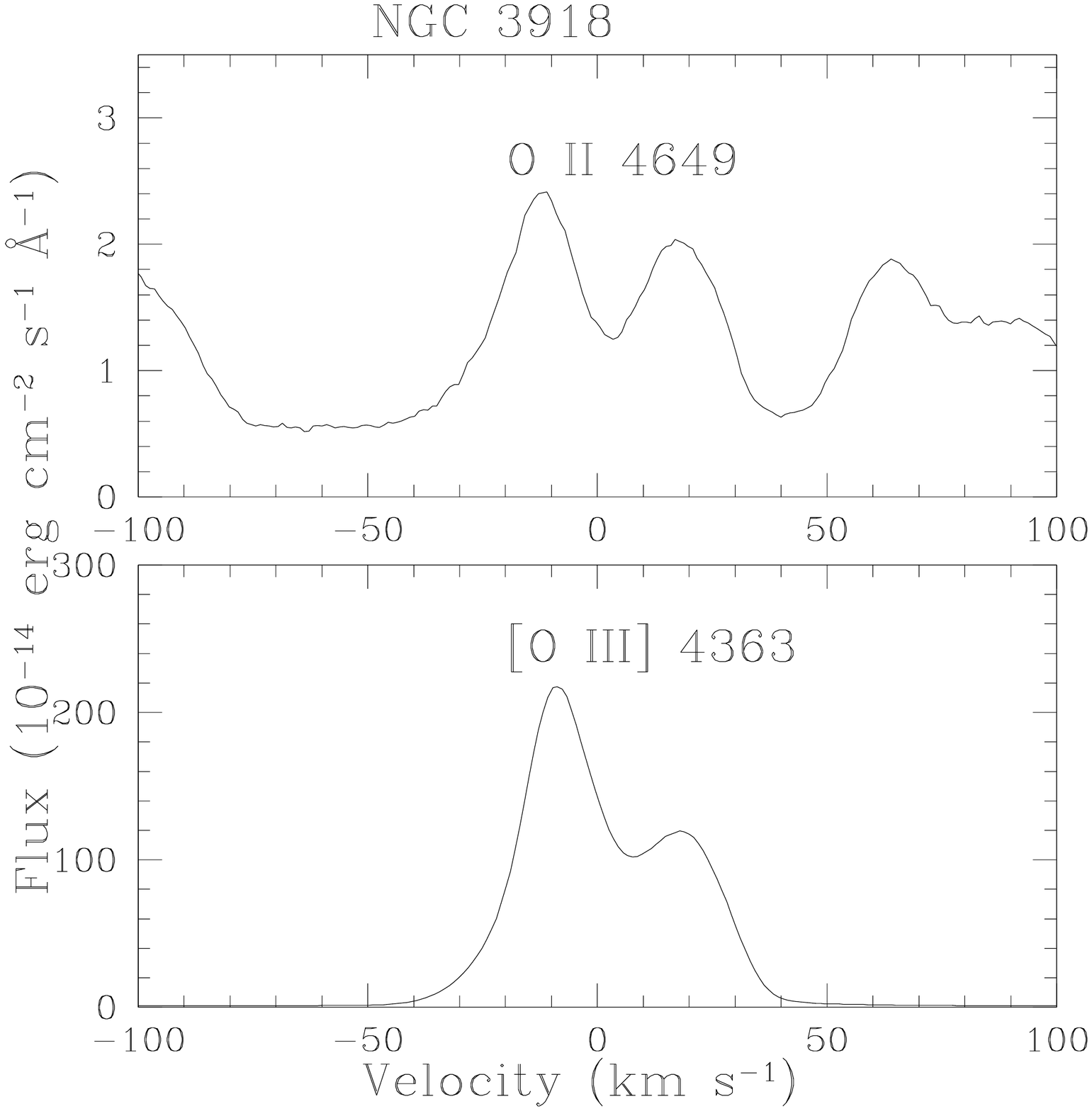}
\includegraphics[scale=0.20]{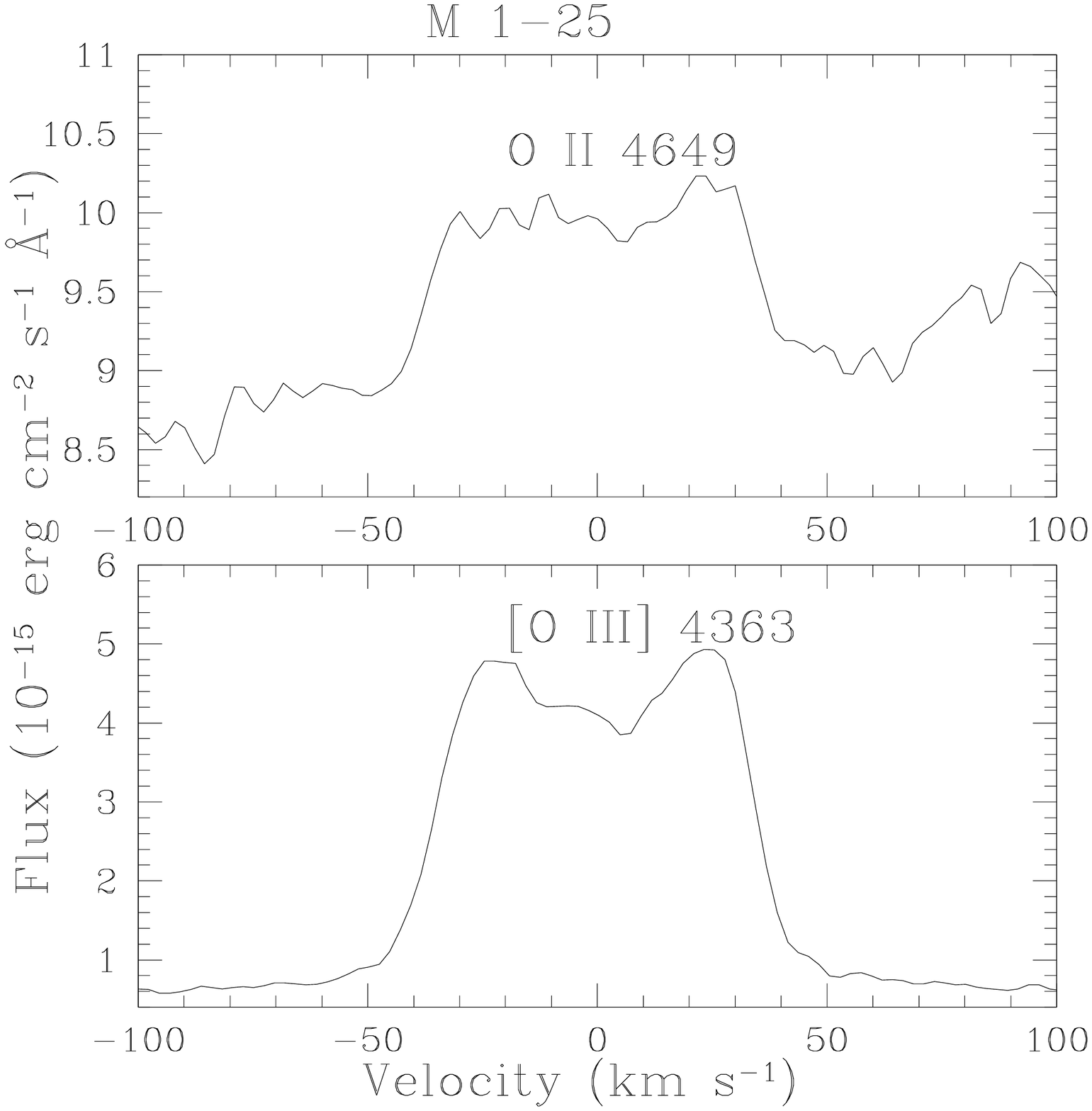}
\includegraphics[scale=0.20]{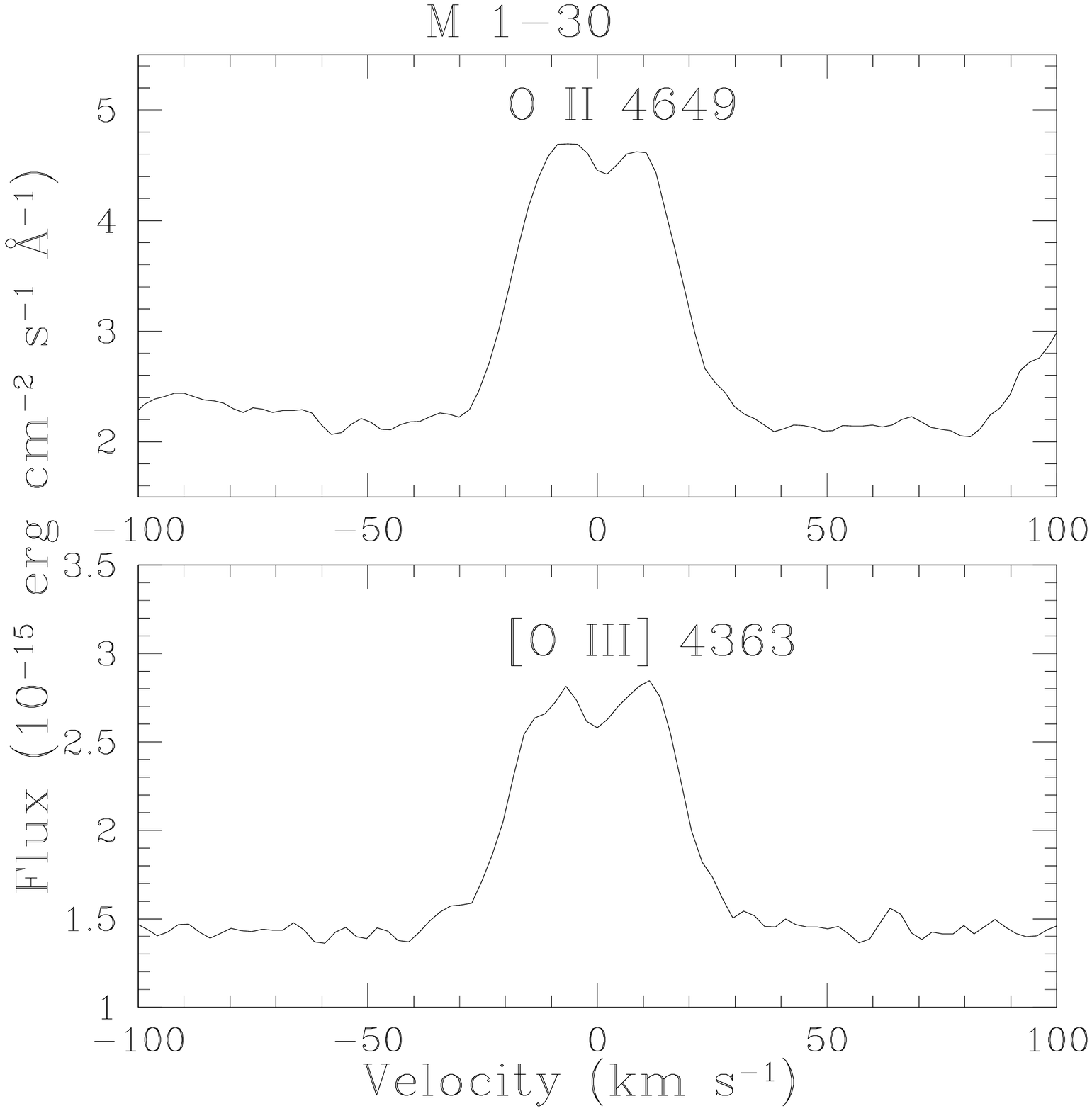}
\includegraphics[scale=0.20]{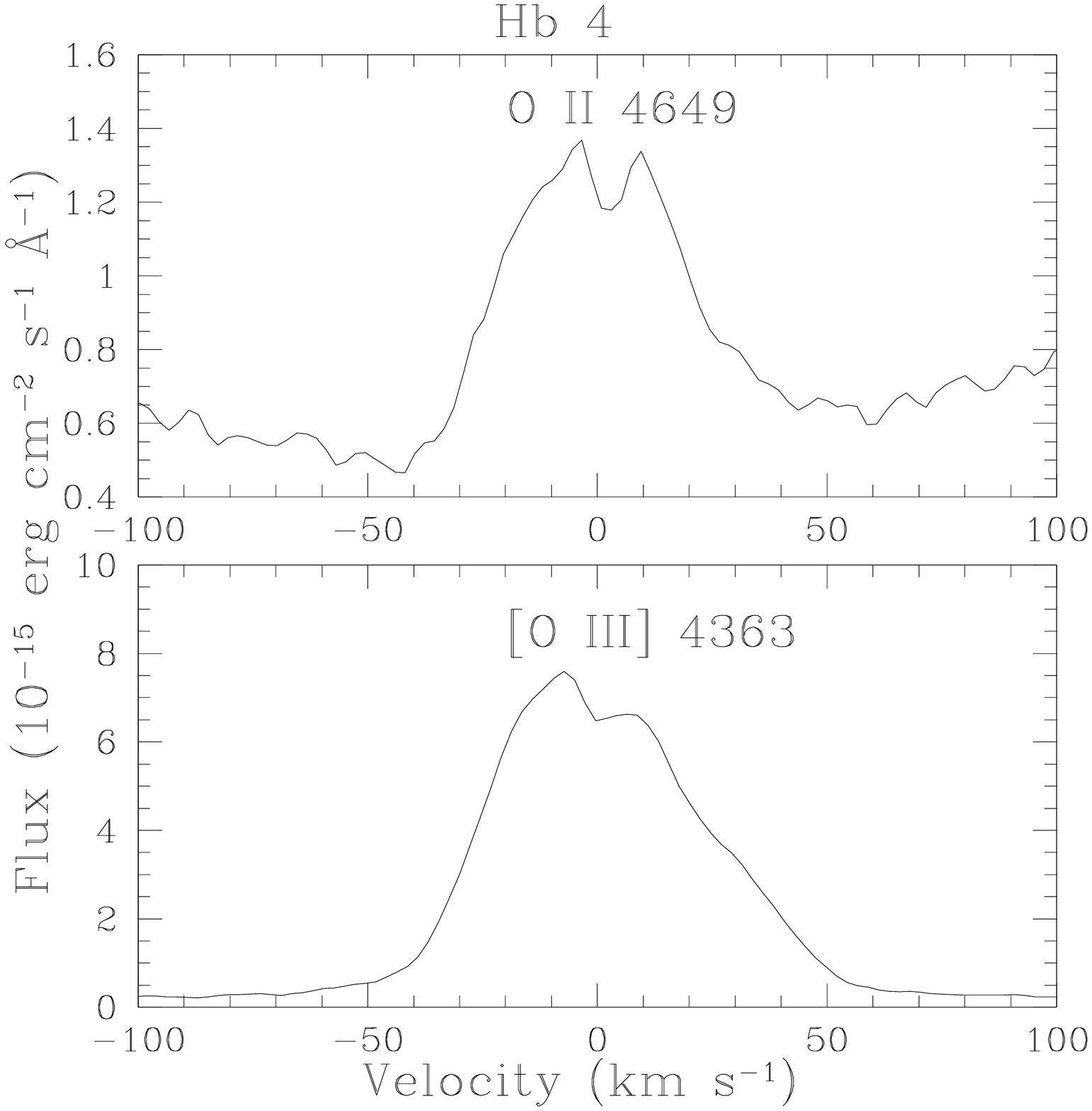}
\includegraphics[scale=0.20]{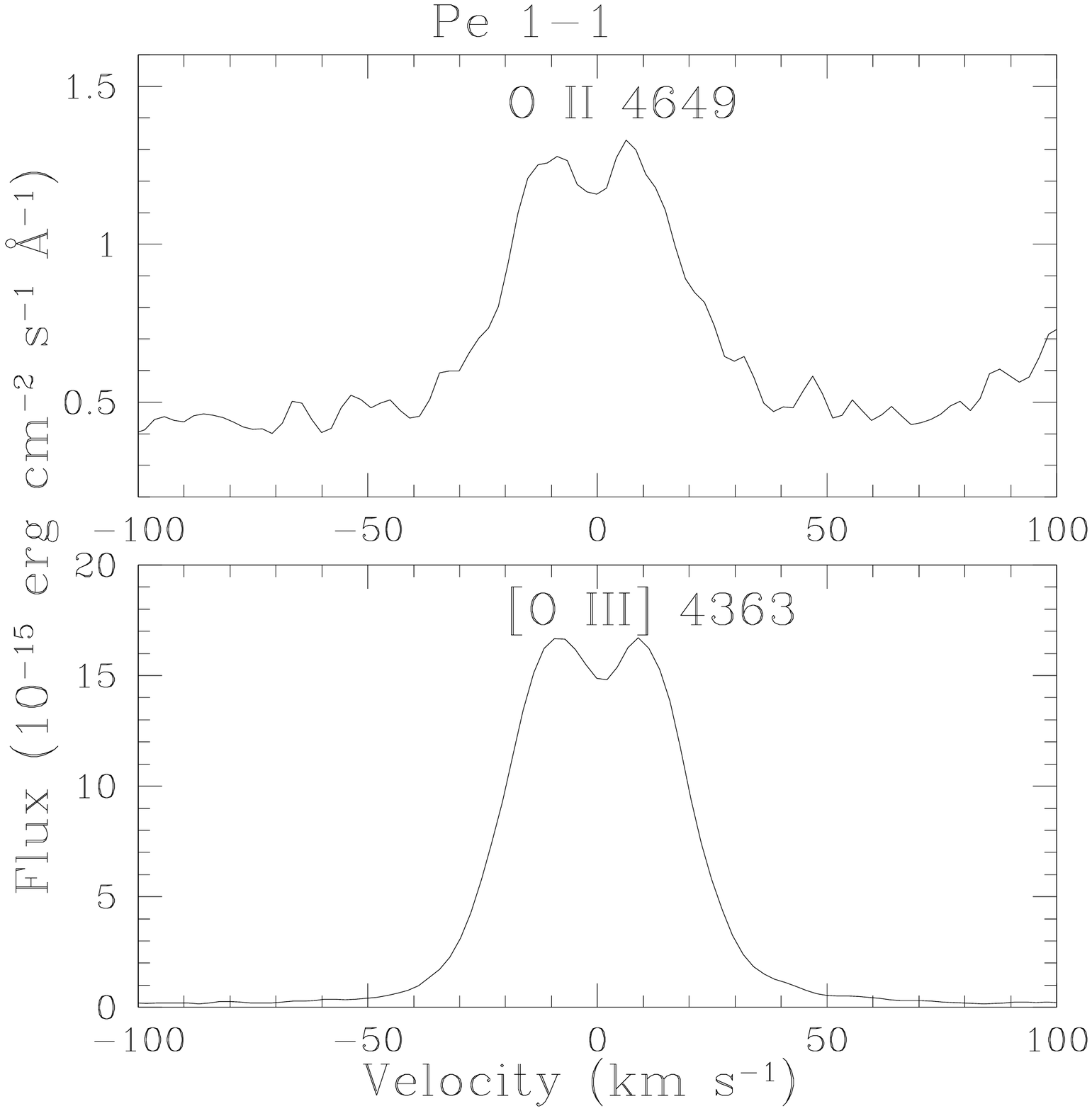}
\includegraphics[scale=0.20]{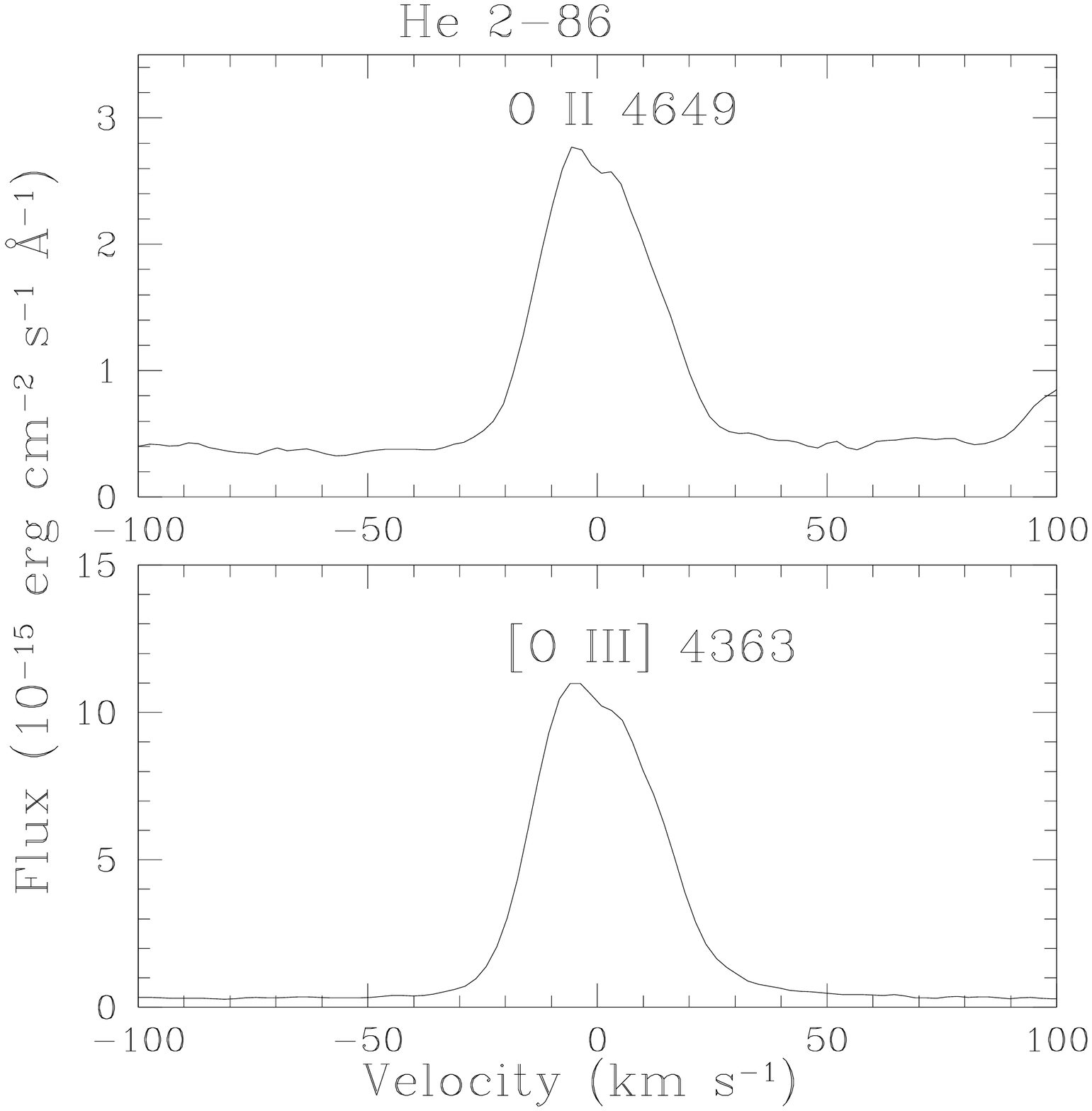}
\includegraphics[scale=0.20]{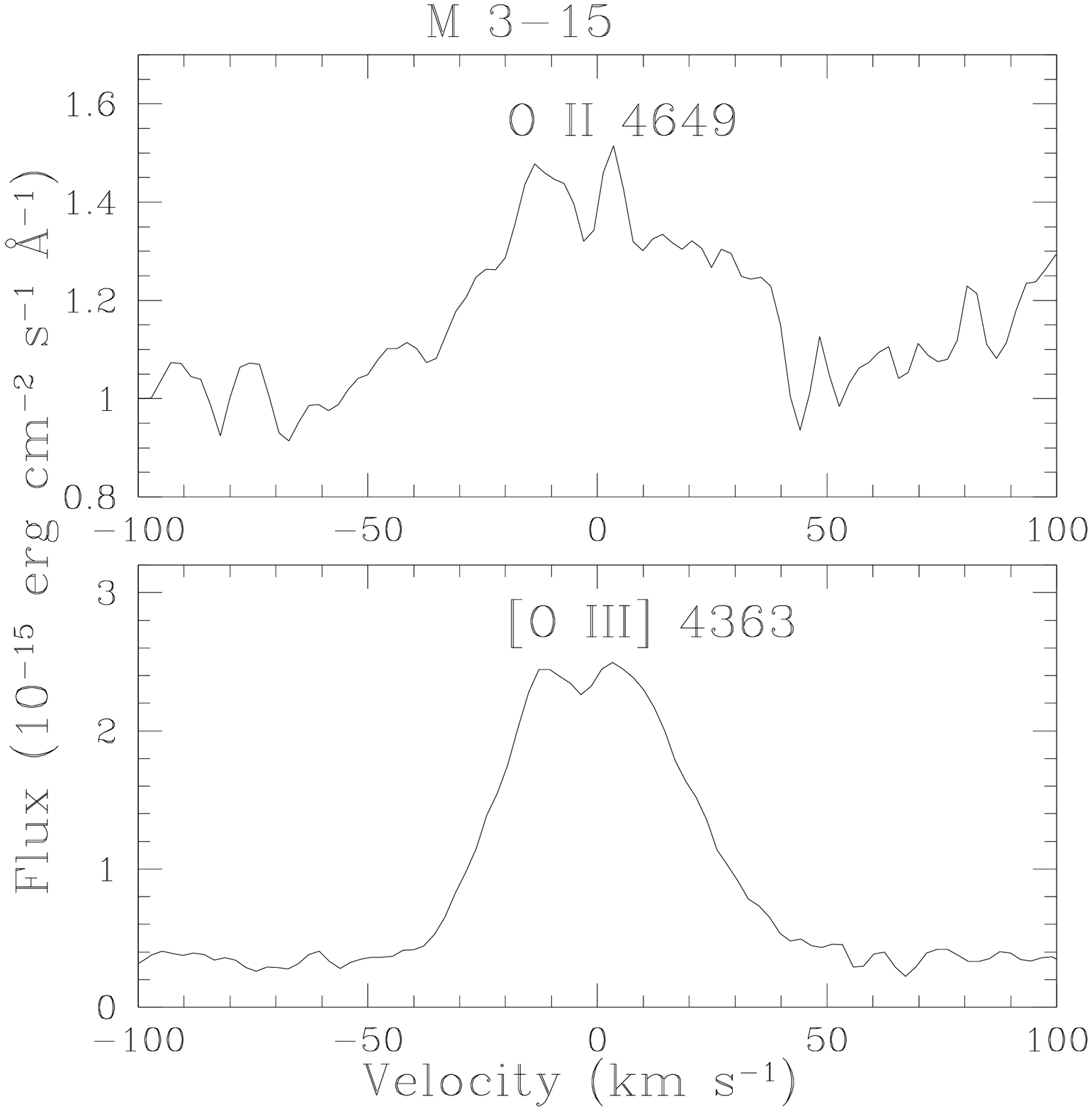}
\includegraphics[scale=0.20]{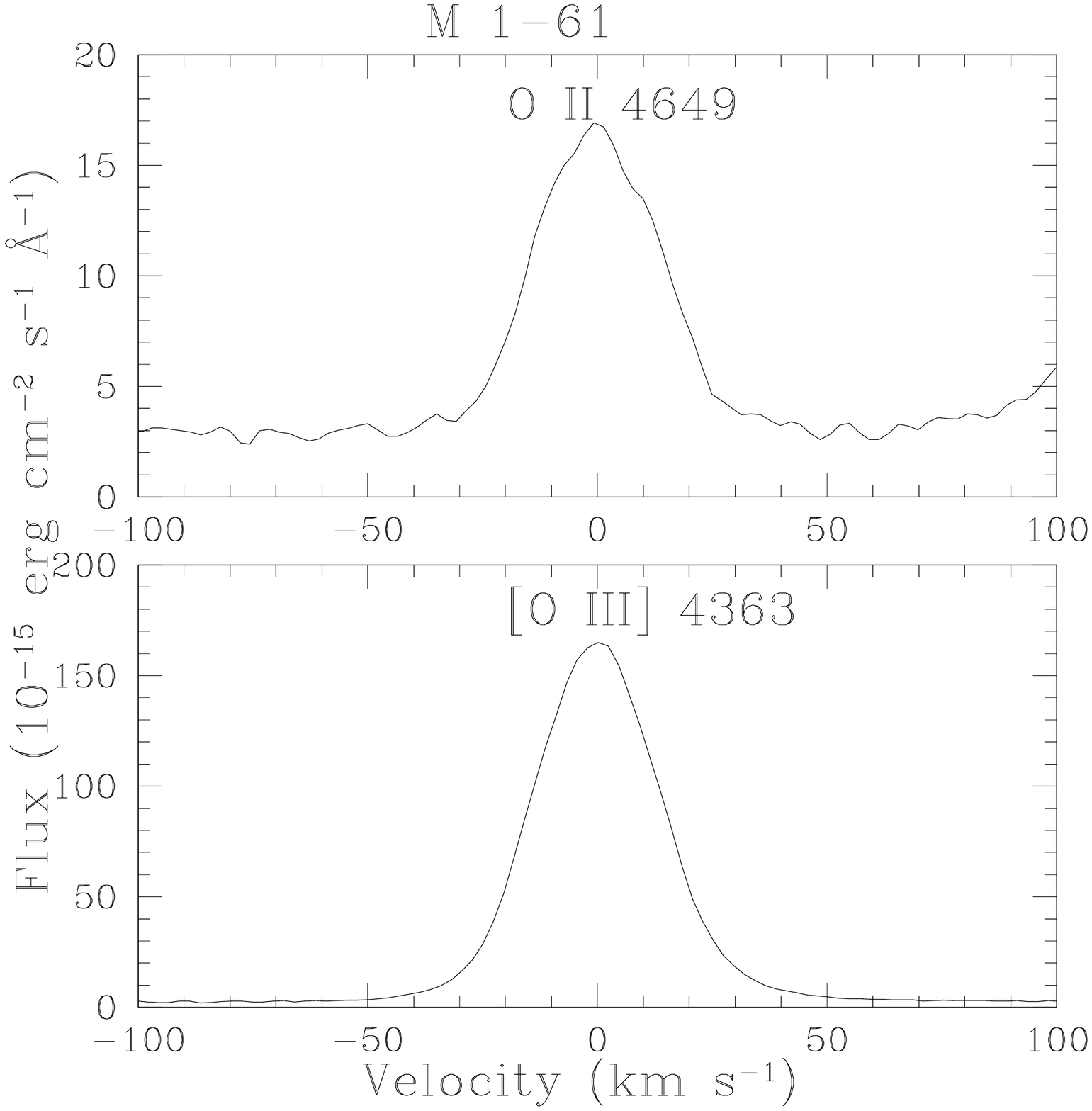}
\includegraphics[scale=0.20]{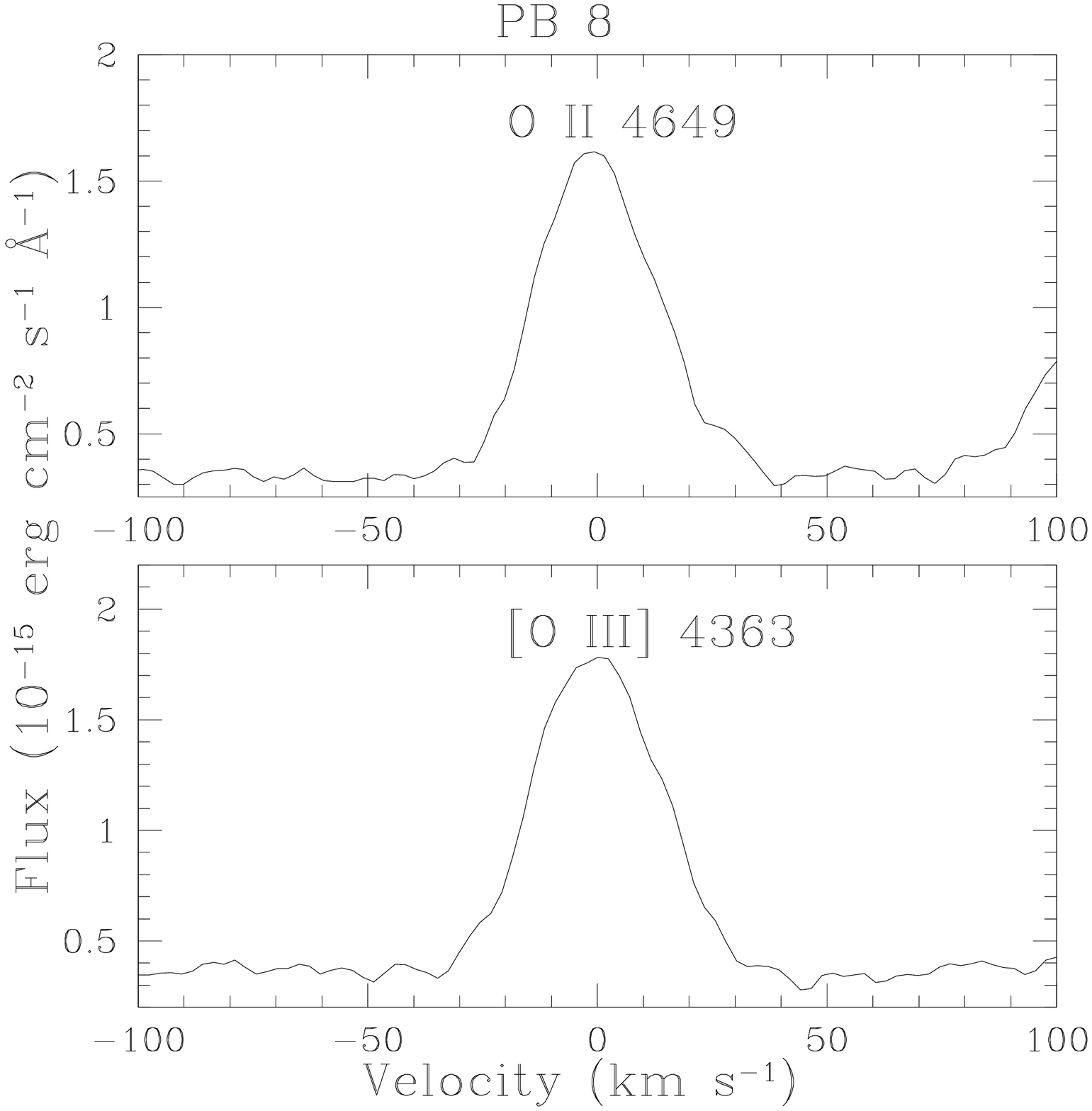}
\includegraphics[scale=0.20]{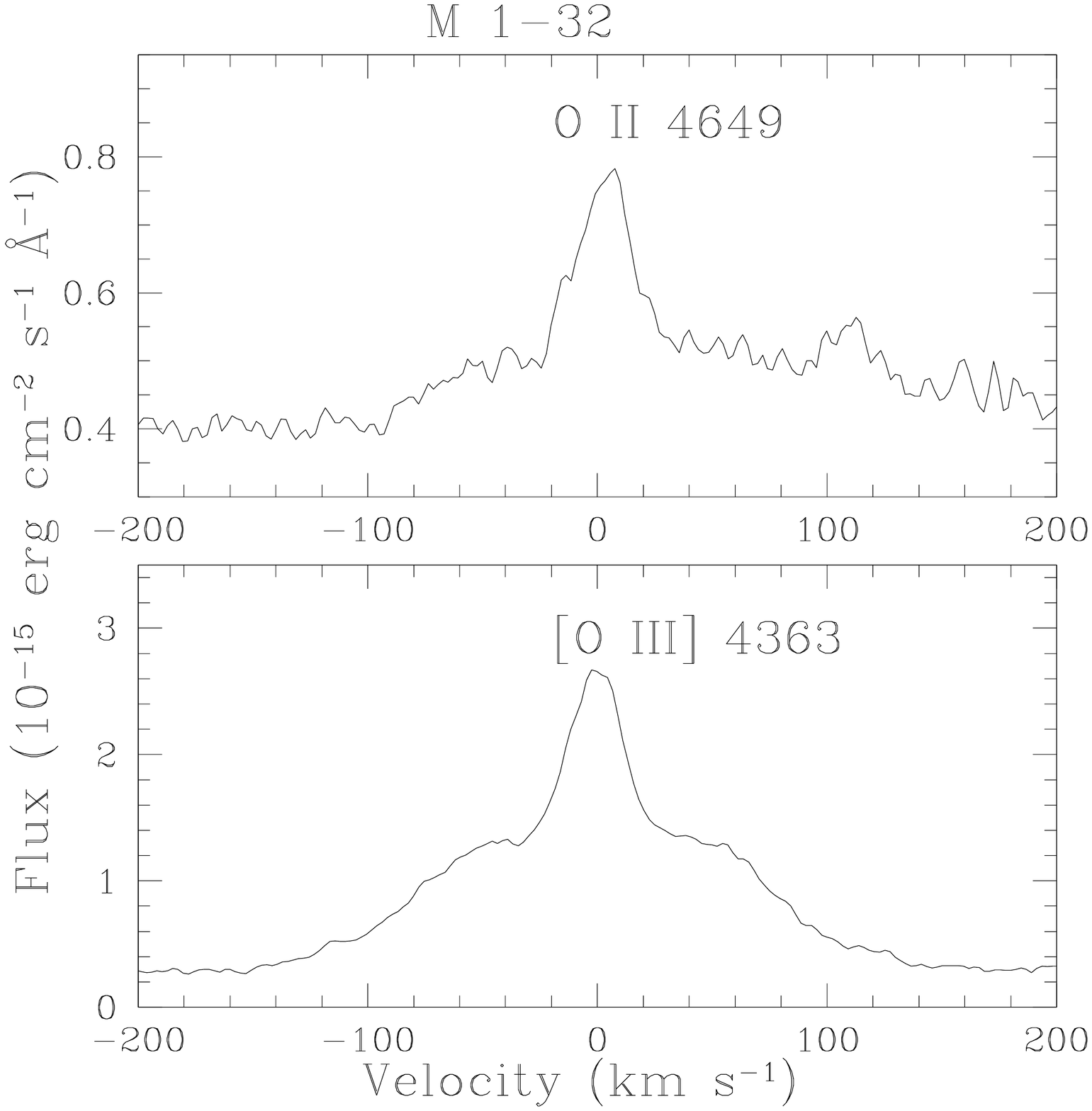}
\end{center}
\end{figure*}

Due to the high spectral resolution of our data (MIKE has R $\sim$ 28000 for a 1$''$ slit width) in most cases the lines are split  showing blue and red components coming from the front and back zones of the expanding shell (e.g., PC\,14, Cn\,1-5, NGC\,2867). In a few cases the lines appear as a single component or show evidence of several overlapped components (e.g., M\,1-61 and PB\,8).  In Fig. \ref{fig:profiles} we show the profiles of the auroral CEL [\ion{O}{iii}] $\lambda$4363  and the ORL \ion{O}{ii} $\lambda$4649, for the objects in our sample. 

 The  line profiles were measured with IRAF {\it splot} routine by fitting a Gaussian profile to the single  lines.  In the cases where the lines are split close, two Gaussian profiles were adjusted in order to deblend the components. Thus, the observed central wavelength and the Gaussian Full Width at Half Maximum ({\it FWHM})  were determined for several ORLs and CELs in each object.  In every case line widths were corrected for the instrumental width, considering that  they add in quadrature. Instrumental widths were adopted from the instrumental spectral resolution of  MIKE spectrograph, $\Delta \lambda \sim$ 0.15 \AA \ in the blue wavelength range and $\sim$ 0.22 \AA \ in the red  one. Thermal widths were not corrected, therefore line widths include thermal broadening and also, turbulence and  physical structure of the nebula along the line of view. 
 
 Results are presented in Table \ref{tab:data} (available on-line, here we present only an example)  where column 1 gives ion ID, column 2 the observed central wavelength, column 3 the rest wavelength, column 4 presents the measured FWHM of the line, column 5 the observed radial velocity and column 6, the expansion velocity measured for the line. The average expansion velocities, V$_{exp}$, and the average {\it FWHM} (for the blue components in case of split lines),  from ORLs or CELs   for each  ion are at the end of the ion list. The  associated uncertainty corresponds to the dispersion of all the CEL or ORL values for each ion, at one sigma.  The formal uncertainty due to the Gaussian fit of lines is not included in this value. In general this uncertainty is small, depending of the rms value of the adjacent continuum and the signal-to-noise of the line which, most of the times, is better  than $\sim$ 10. The comparison of V$_{exp}$ for different lines of the same ion provides the real uncertainty. 

The same procedure was applied to data from the literature for NGC\,7009 and NGC\,3918. The values for NGC\,7009 were extracted from \citet{richer:13} and the values for NGC\,3918 from the deep high-resolution spectrum presented by \citet{garcia:15}, obtained with the UVES spectrograph  at  the ESO- VLT.  The line widths in this spectrum were corrected for the instrumental width of UVES.

In the case of objects with split lines, V$_{exp}$  were computed as half the difference between the blue and red components of lines which is a good measurement of the expansion velocity 
if the nebula is well resolved \citep{gesicki:00}.  For the few objects with single lines, the expansion velocity V$_{exp}$ was assumed to be half the {\it FWHM}, which is a good approach for a unresolved or partially resolved nebula \citep{gesicki:00}, although in this case the {\it FWHM} includes the expansion as well as  thermal broadening, turbulence, nebular structure, etc.  In most cases we are measuring the expansion velocity of the   bright zone of the shell  and not the true expansion velocity which for a multishell nebula could happen in a fainter outer zone \citep[see][for a deep discussion of the expansion velocity structure]{schonberner:14}. As our intention is to compare the same parameter for ORLs and CELs, this is a valid approach.

 For each ion V$_{exp}$ were measured for all available CELs or ORLs,  and the average values  were calculated for CELs and ORLs separately. Average V$_{exp}$ derived this way are presented in graphic form as a function of the ionisation potential (IP) for the different ions, for each object (Figs.~\ref{fig:graphics-1} and \ref{fig:graphics-2}). The errors in velocity correspond to  1 $\sigma$ deviation of the average (except for the cases where only one line was measured, in which case $\sigma$ = 1 km s$^{-1}$ was adopted). For these graphs, we have chosen the IP because, for a plasma in ionisation equilibrium there exists a stratification of the ionisation structure within the nebula, where it is found that highly ionised species are near the central star while the low ionised species are farther away. Therefore  in some way the IP  represents the distance of the ion location relative to the central star \citep[see e.g., Fig 2.6 by ][]{osterbrock:06}.

\begin{table}
\caption{(Full table available on-line) Measured data for Cn\,1-5 \label{tab:data}}
\begin{tabular}{lllcrr}
\hline 
\hline \\		   
    Ion   &$\lambda_{obs}$ & $\lambda_0$& ${\sc FWHM}$$^a$ &  V$_{rad}~~~$ &   V$_{exp}$  ~~~\\
      &(\AA) & (\AA) & (\AA) & km s$^{-1}$ &   km s$^{-1}$ \\
\hline
 $[{\rm Ar~III}]$ &  7135.82 & 7135.78   & 0.63   & 26.94  &  25.32 \\
 $[{\rm Ar ~III}]$ &  7137.02 & 7135.78   & 0.72    \\
 $[{\rm Ar~III}]$ &  7751.18 & 7751.10   & 0.69  &  28.38  &  25.24 \\
 $[{\rm Ar~III}]$ &  7752.49 & 7751.10   & 0.76 \\
  Average &          &           & 0.66 &       &   25.28 \\
sigma &          &           &   0.03 &       &   0.05\\
\hline
  $[{\rm Ar~IV}]$ & 4711.32 & 4711.37  &  0.49  &  22.96  &  26.04 \\
 $[{\rm Ar~IV}]$ & 4712.14 & 4711.37  &  0.19 \\
  $[{\rm Ar~IV}]$ & 4740.22 & 4740.17  &  0.51  &  29.49  &  26.04 \\
  $[{\rm Ar~IV}]$ & 4741.05 & 4740.17  &  0.46 \\
  Average &         &          &      0.50  &         &  26.04   \\
sigma&          &           &    0.01  &       &   0.10\\
\hline
 $[{\rm Cl~II}]$ & 8578.73 & 8578.70  &  0.74  &  26.65  &  25.65 \\
 $[{\rm Cl~II}]$ & 8580.20 & 8578.70  &  0.73 \\
  $[{\rm Cl~II}]$ & 9123.63 & 9123.60  &  0.80  &  26.28  &  25.20 \\
 $[{\rm Cl~II}]$ & 9125.17 & 9123.60  &  0.78 \\
  Average &         &                   &     0.77 &            & 25.42 \\
sigma  &          &                &   0.04  &       &   0.32\\
\hline 
\multicolumn{6}{l}{$^a$ {\it FWHM} of the blue component is listed.}
\end{tabular}
\end{table}

\section{ Behaviour of expansion velocities}

\begin{figure*}
  %\begin{center}
{\includegraphics[scale=0.45]{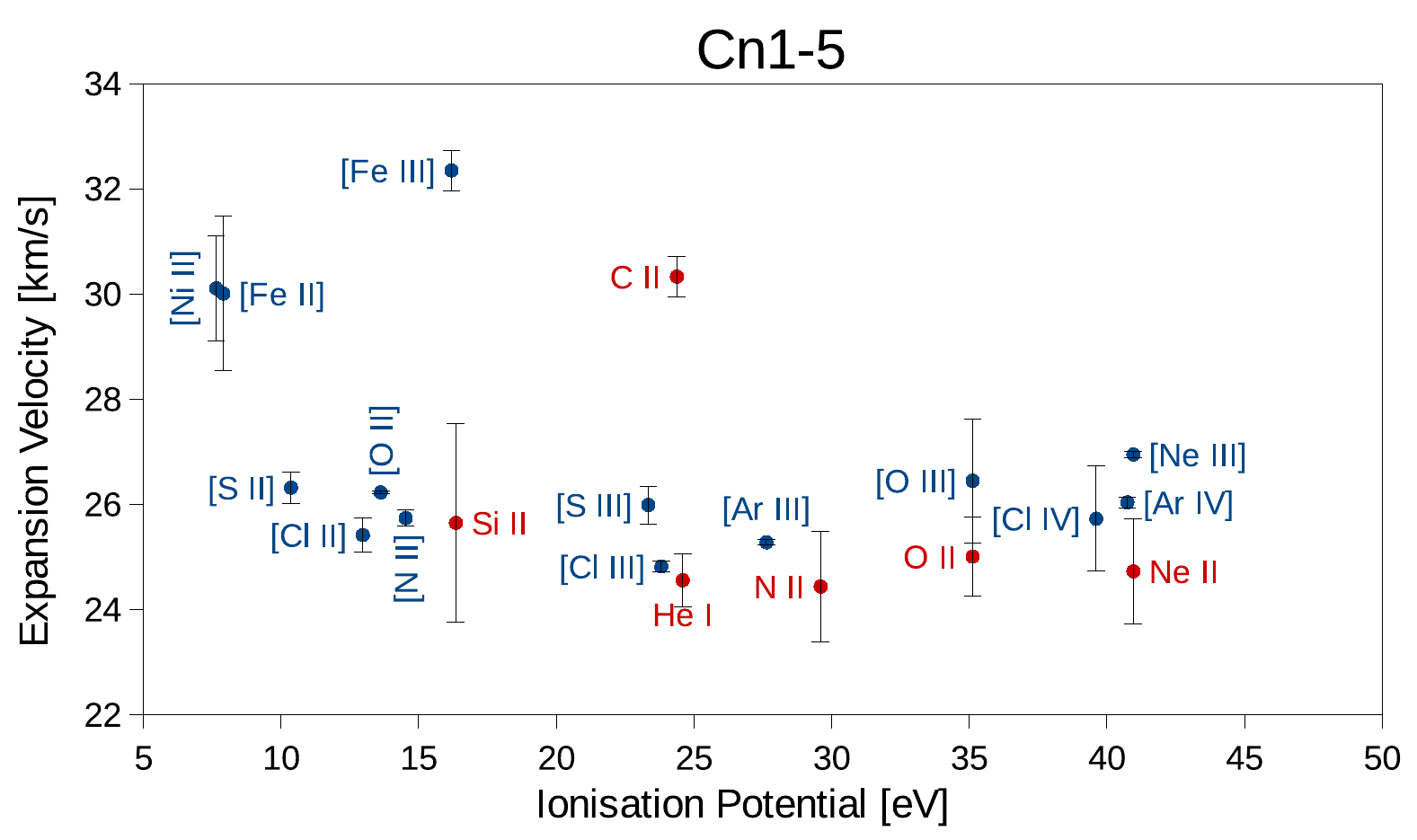}}
{\includegraphics[scale=0.45]{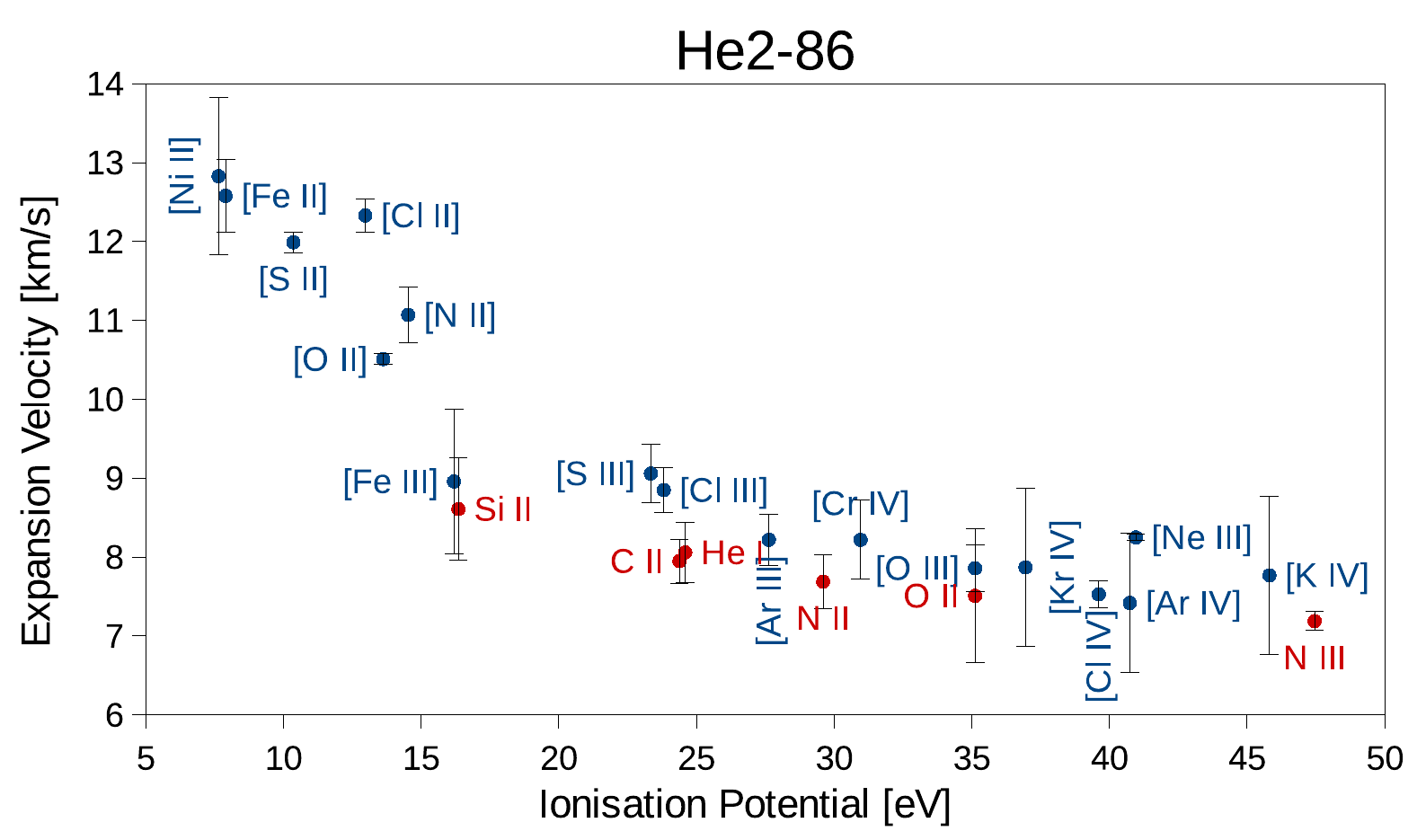}}
{\includegraphics[scale=0.45]{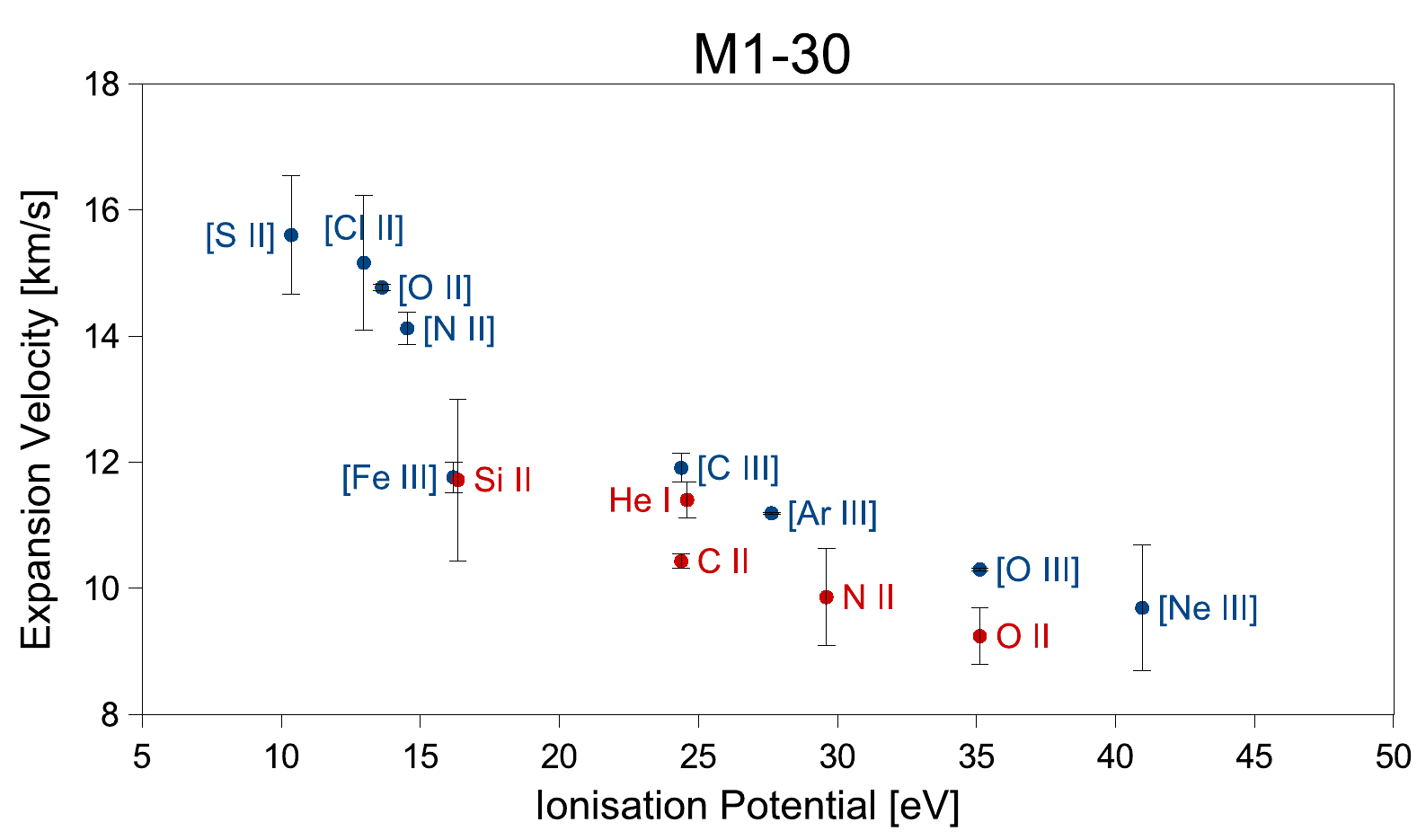}}
{\includegraphics[scale=0.45]{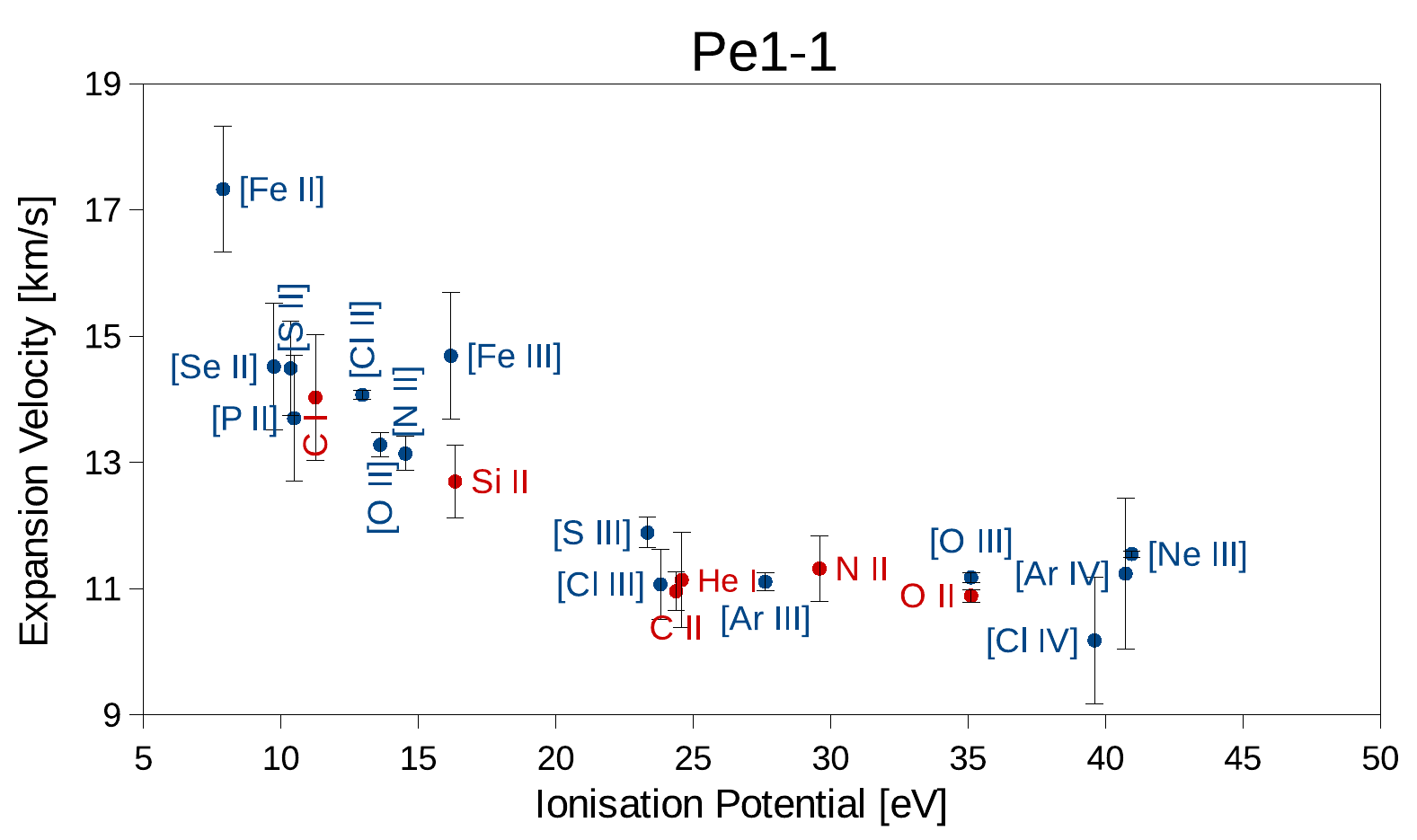}}
{\includegraphics[scale=0.45]{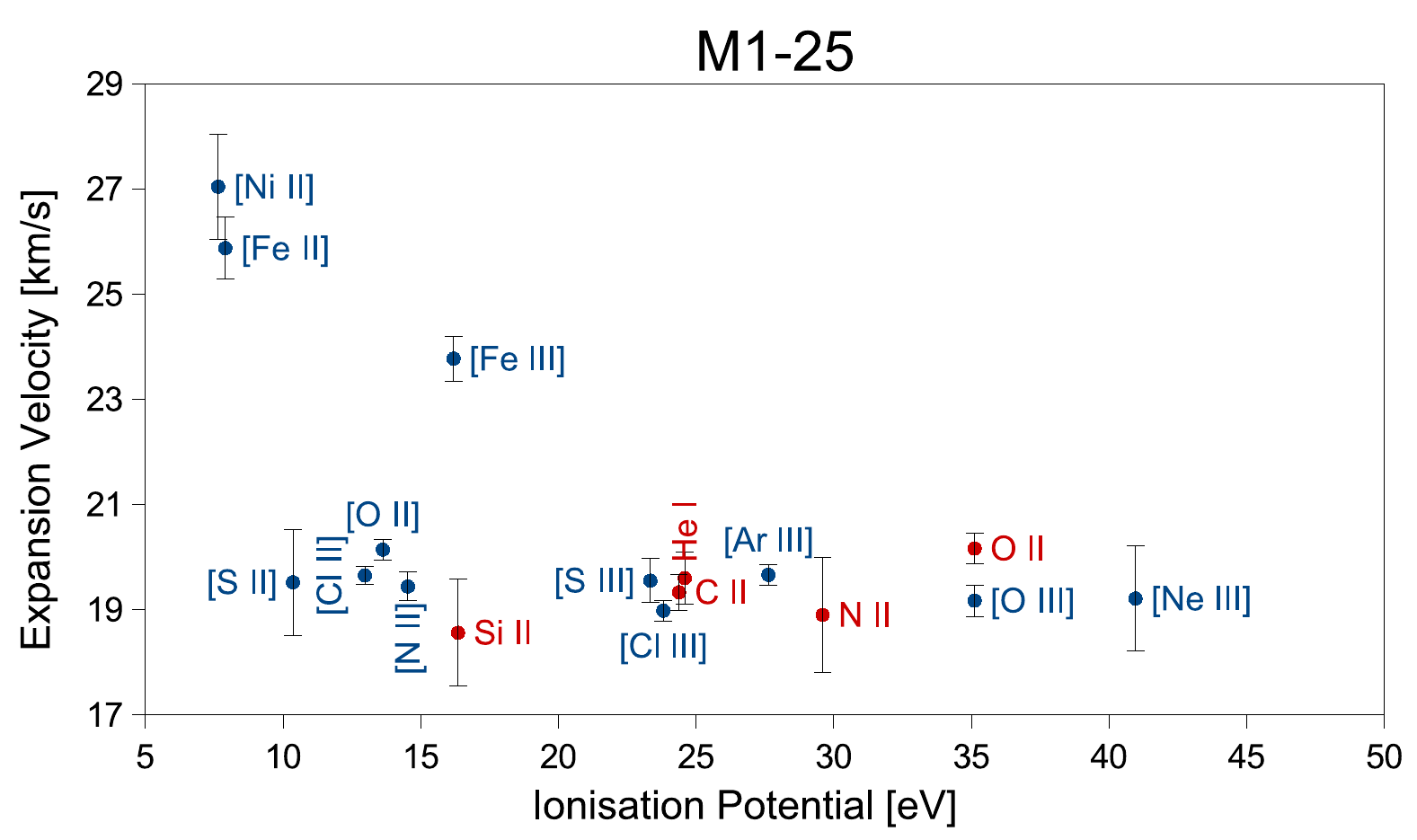}}
{\includegraphics[scale=0.45]{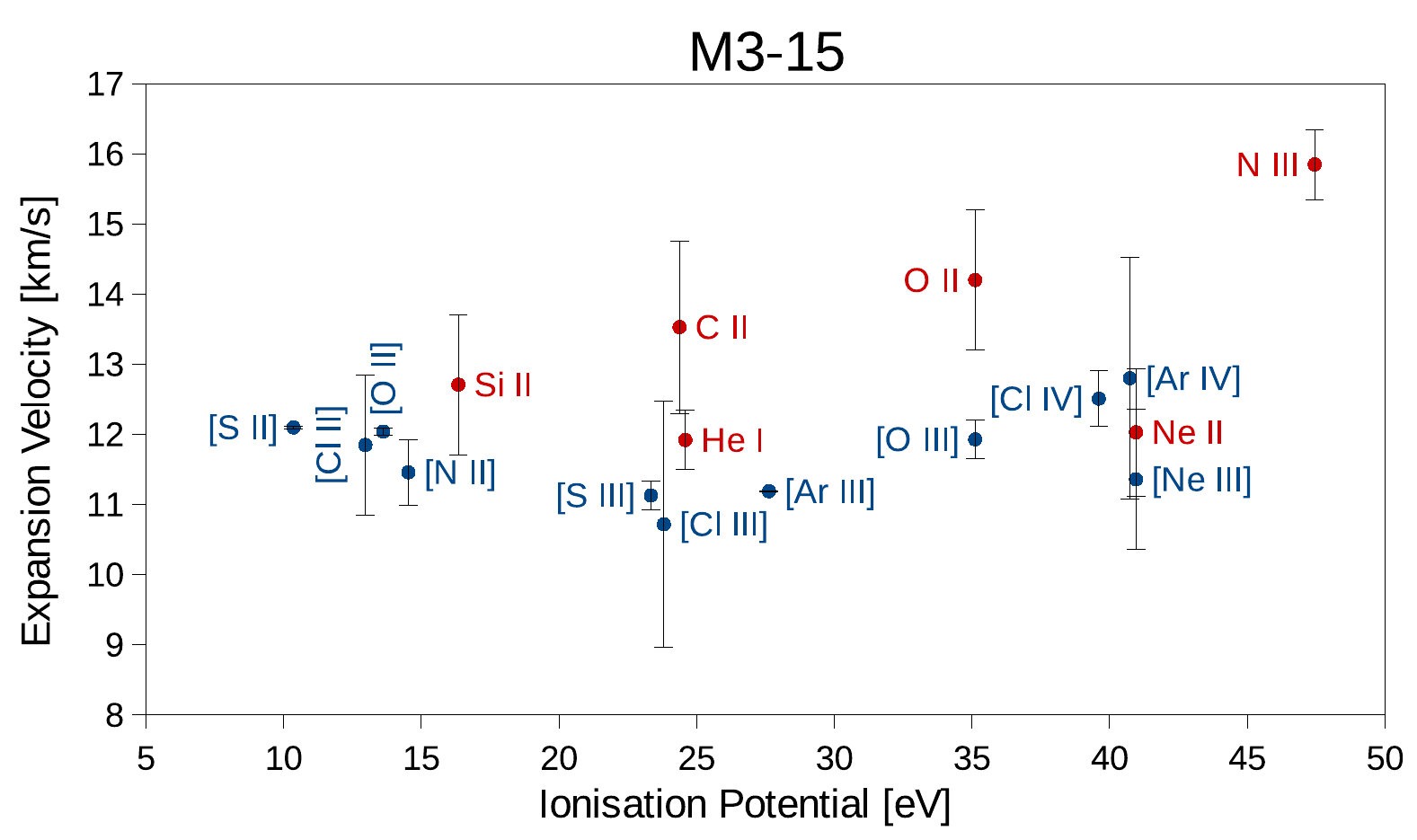}}
{\includegraphics[scale=0.45]{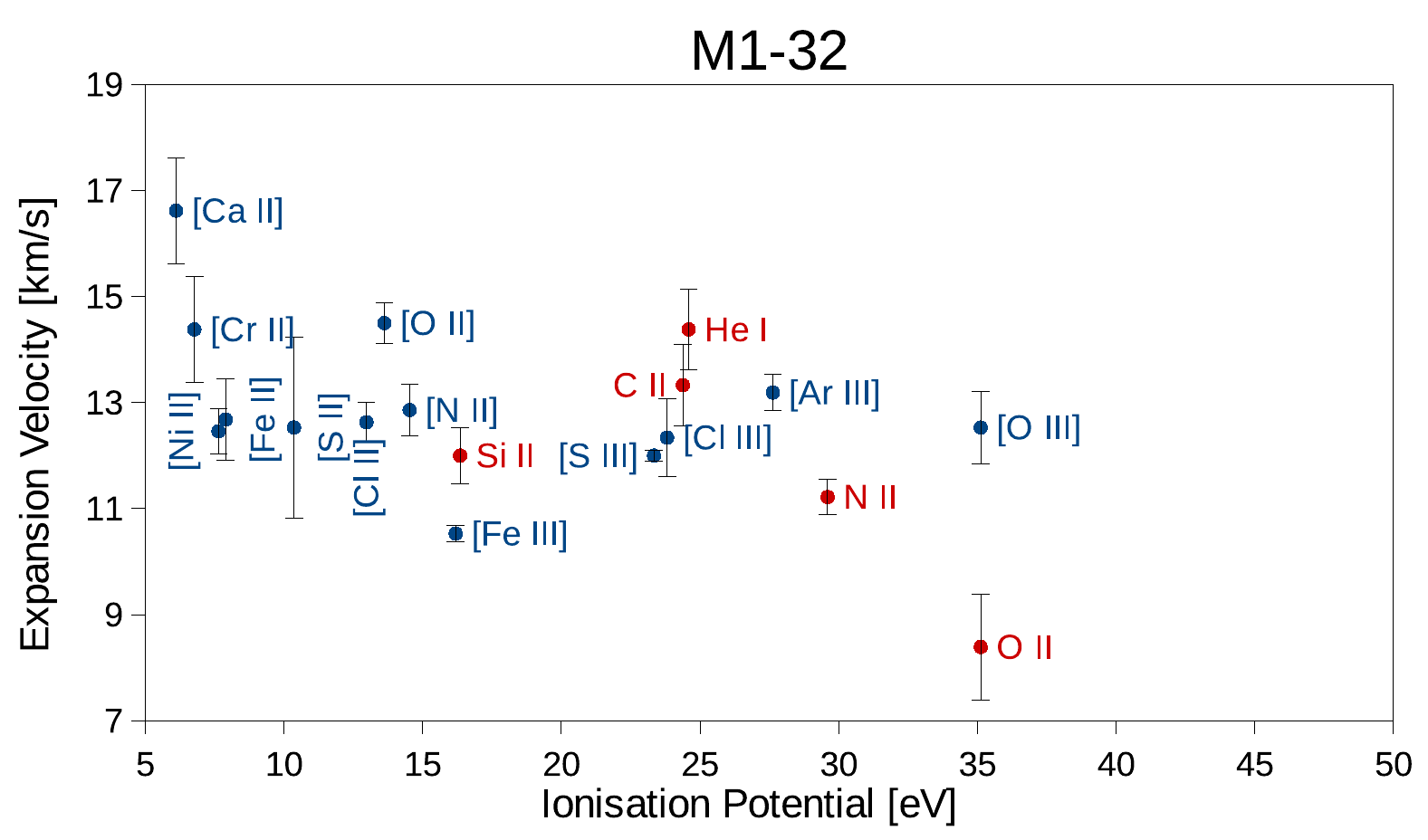}}
{\includegraphics[scale=0.45]{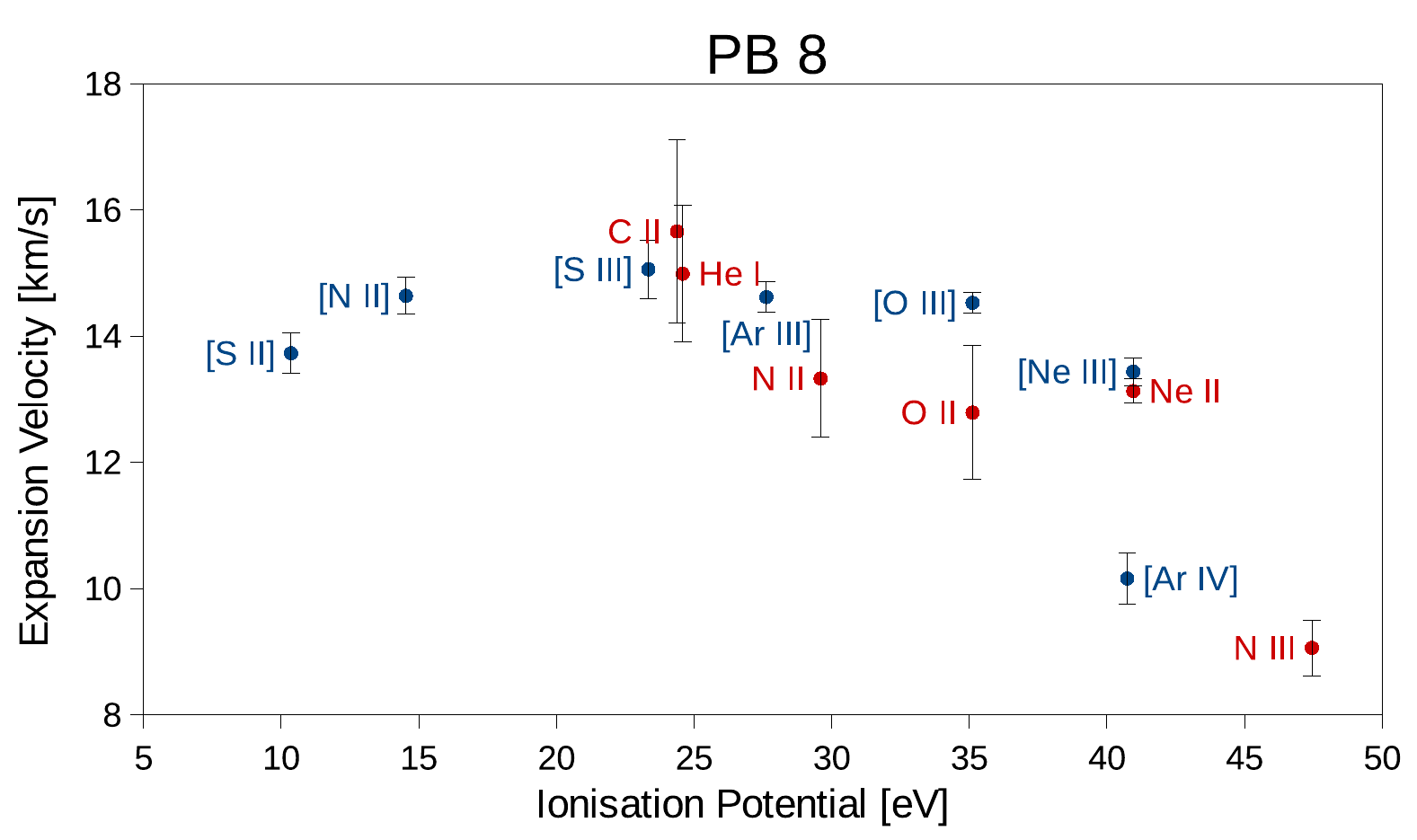}}
 \caption{V$_{exp}$ vs. Ionisation Potential for CELs (blue symbols) and ORLs (red symbols) for low excited PNe. \label{fig:graphics-1}}
%\end{center}
\end{figure*}

\begin{figure*}
  \begin{center}
{\includegraphics[scale=0.45]{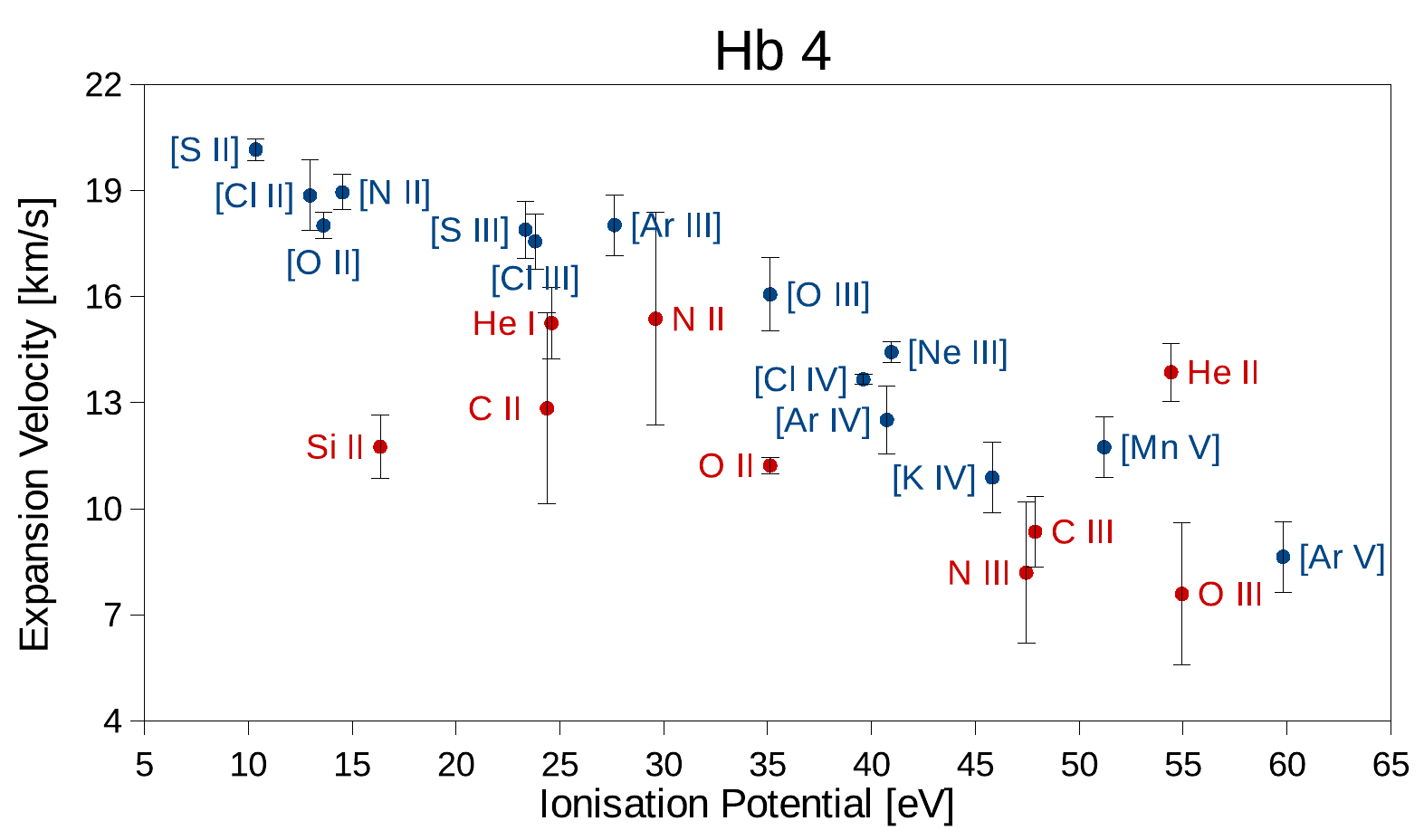}}
{\includegraphics[scale=0.45]{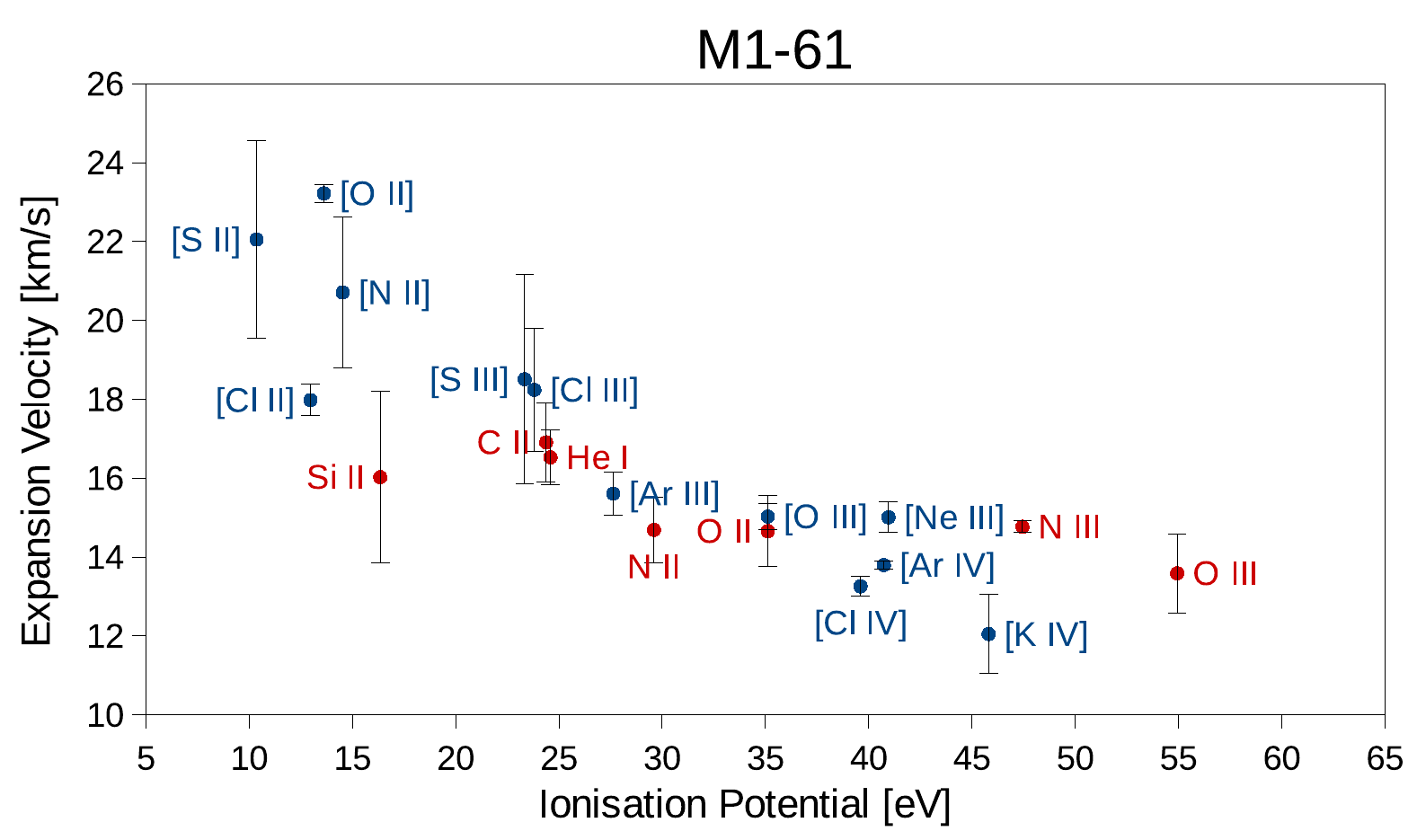}}
{\includegraphics[scale=0.45]{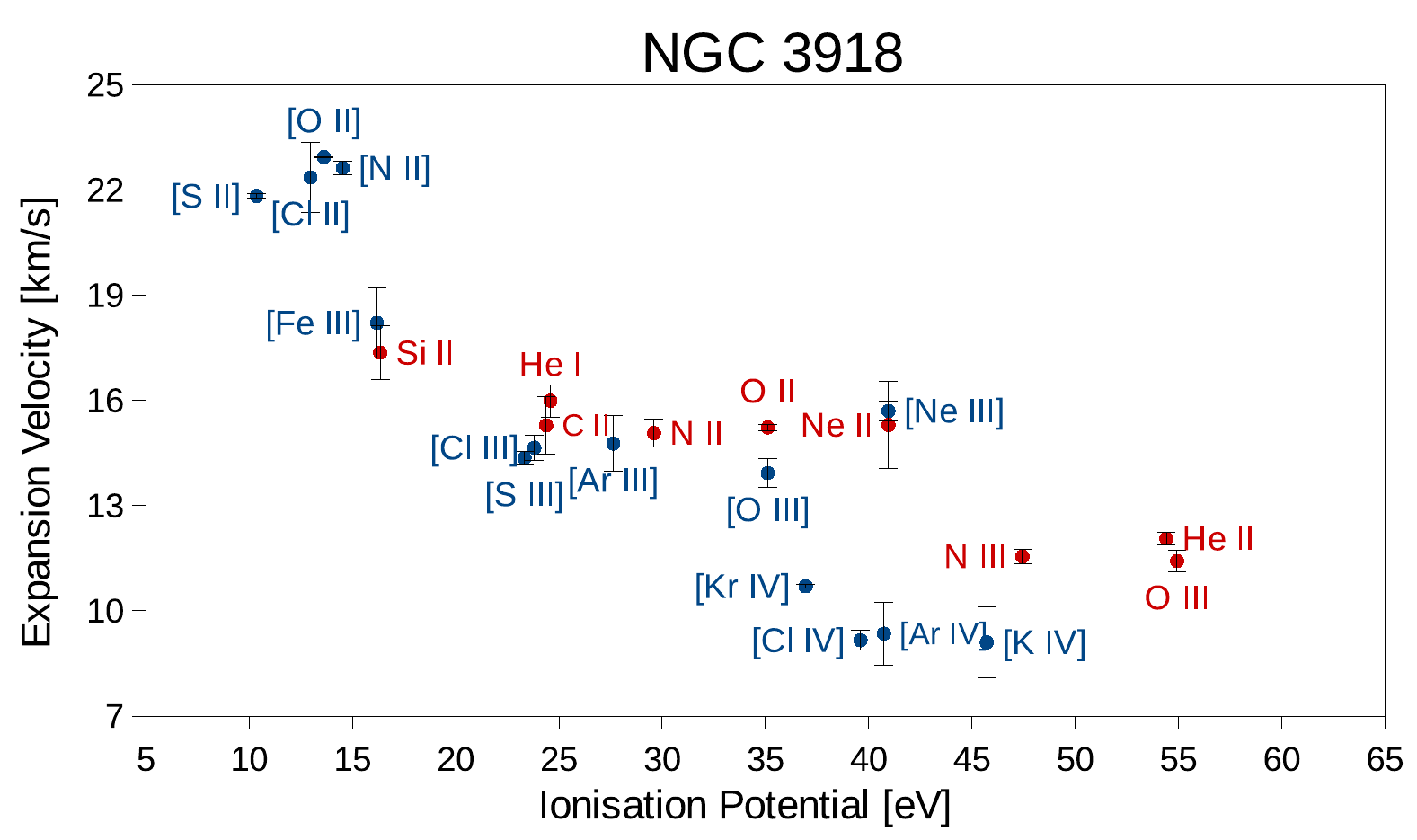}}
{\includegraphics[scale=0.45]{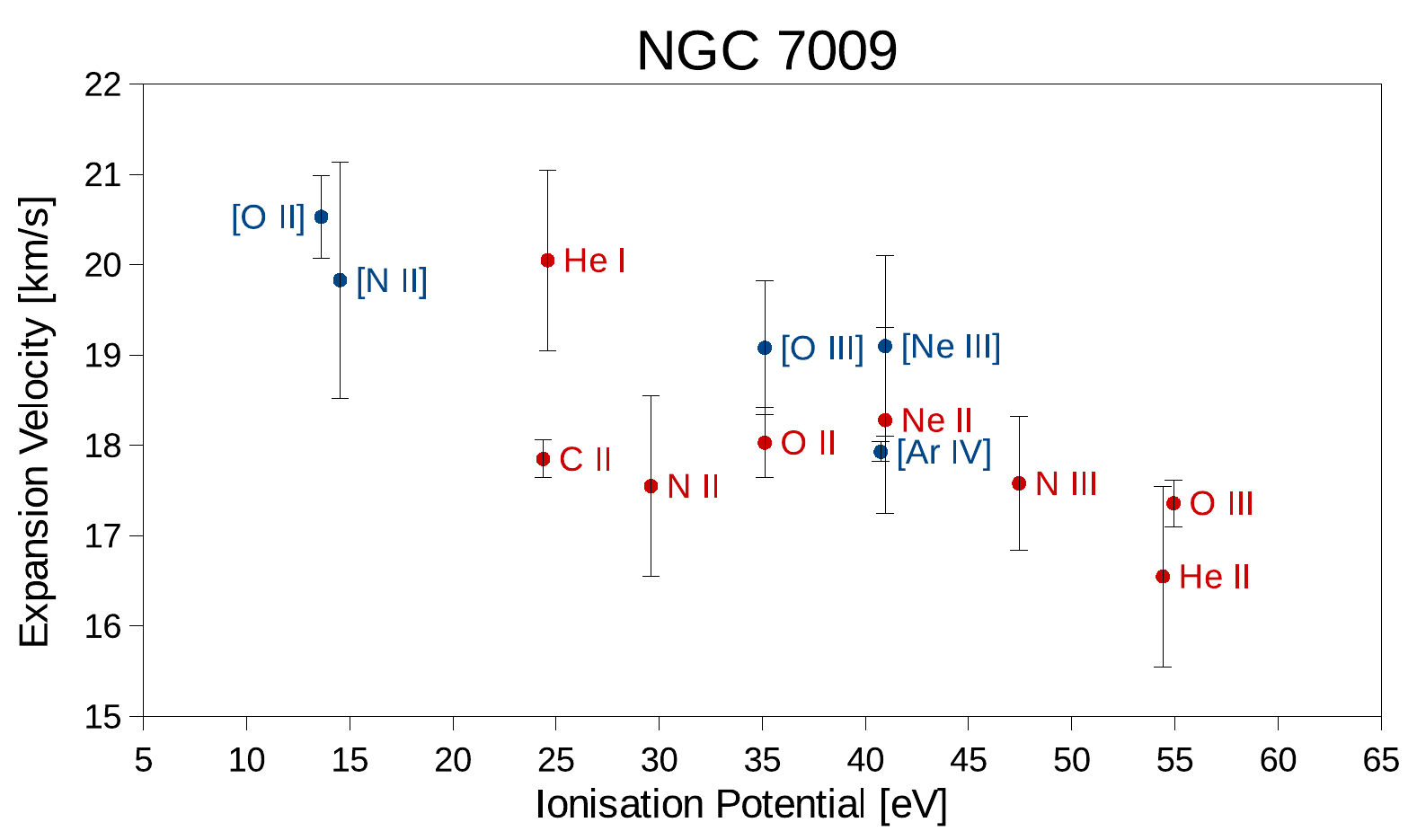}}
{\includegraphics[scale=0.45]{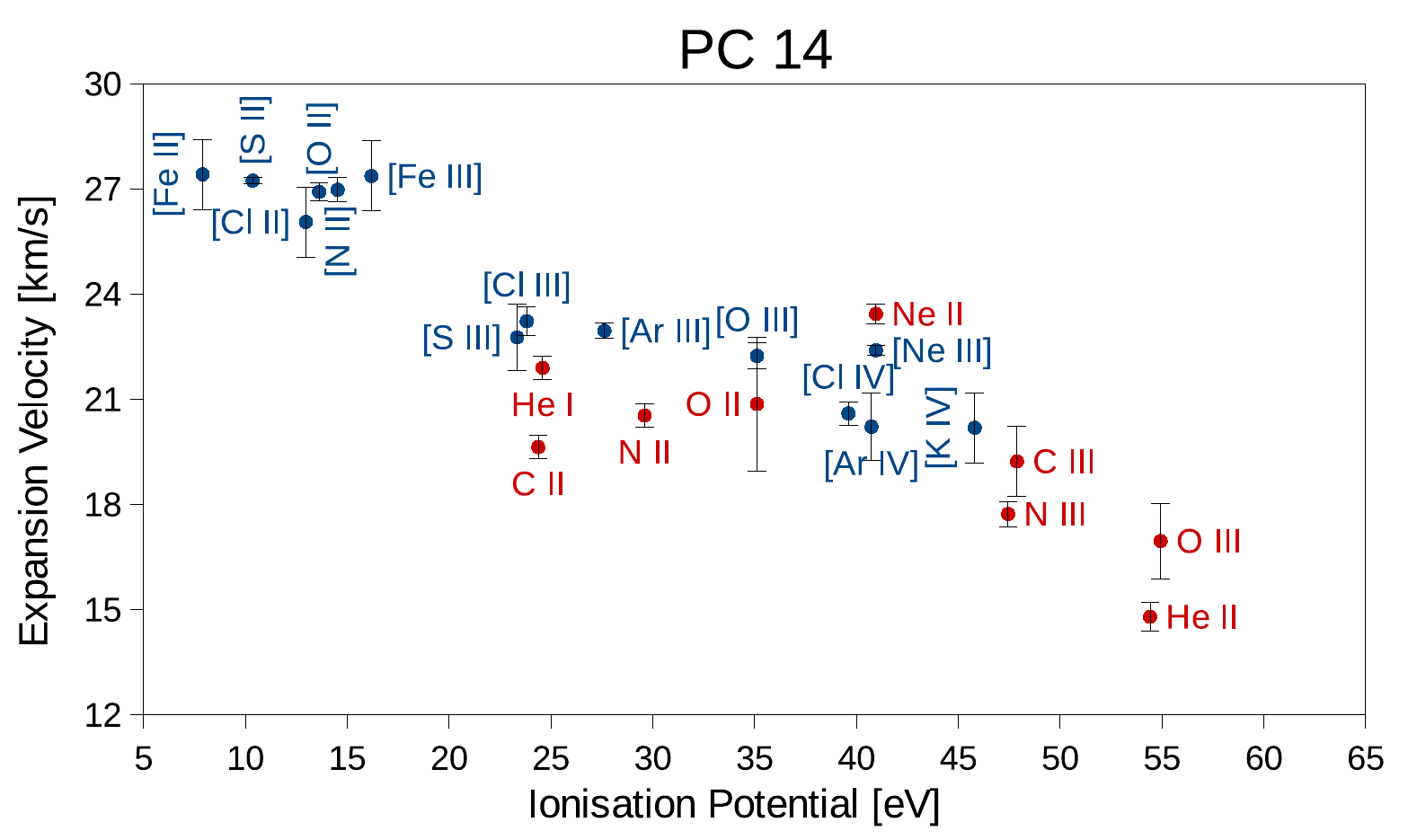}}
{\includegraphics[scale=0.45]{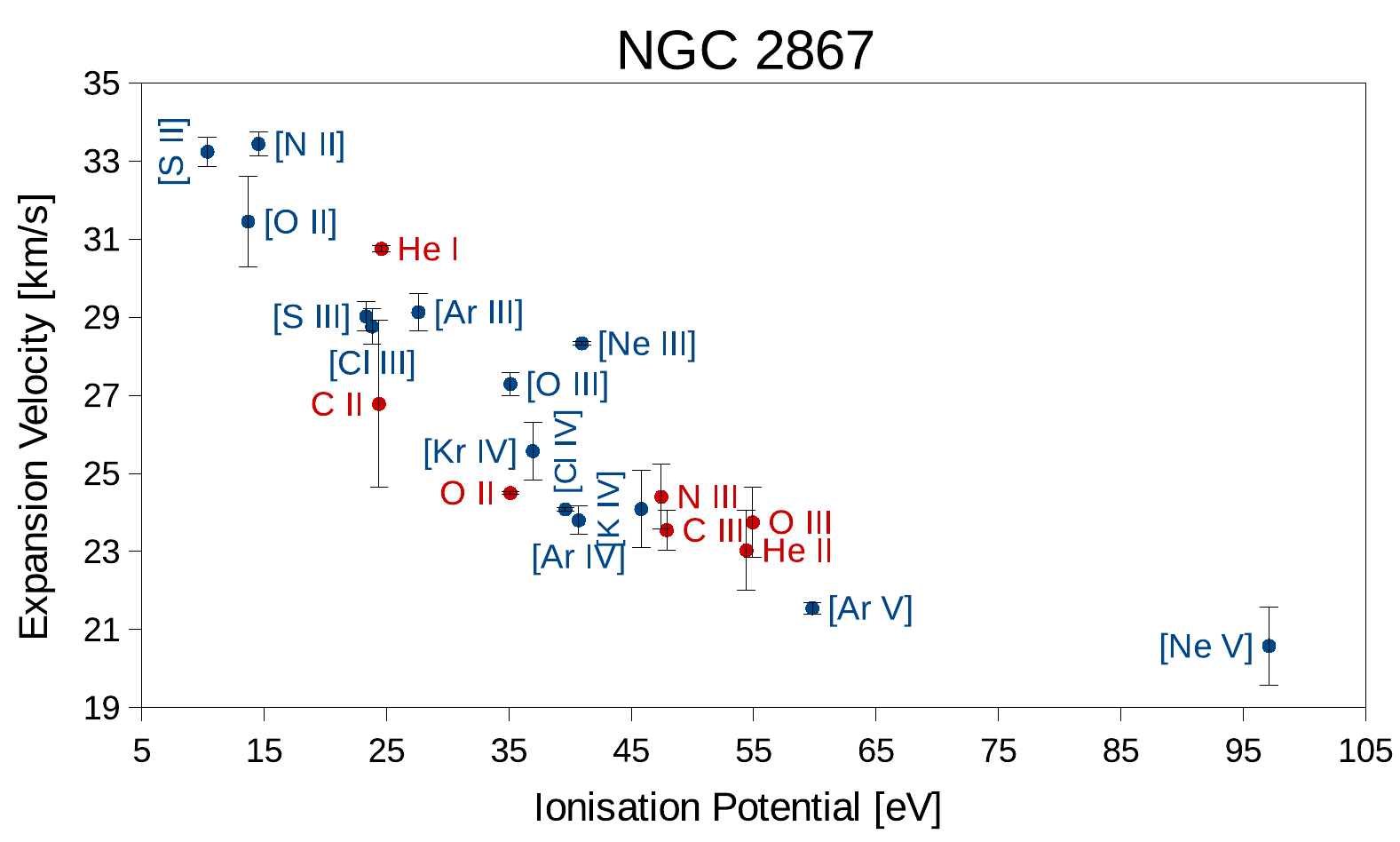}}
  \caption{V$_{exp}$ vs. Ionisation Potential for CELs (blue symbols) and ORLs (red symbols) for highly ionised PNe. \label{fig:graphics-2}}
\end{center}
\end{figure*}

In the graphs presented in Figs. \ref{fig:graphics-1} and \ref{fig:graphics-2} we plot, for all the analysed objects, the expansion velocity of each ion, as derived from the average of its ORLs (in red) and the average of its CELs (in blue), as a function of its IP.  For an easy way of comparing the behaviour of different objects, in Fig. \ref{fig:graphics-1} we include all the PNe whose IP of ions cover up to 50 eV (these are relatively low excitation nebulae) while in Fig. \ref{fig:graphics-2} the highly ionised nebulae are presented. 

In these figures we have not included the velocities  measured for the permitted lines of \ion{N}{i} and \ion{O}{i} (although they are listed in Table \ref{tab:data}) because these lines could be mainly excited by direct starlight, therefore they would not be  {\it bona fide} recombination lines \citep{grandi:75}  and their  emission would be produced outside the ionised nebula. 

On the other hand, \citet{grandi:76}  has also claimed that permitted lines from Si$^+$ may be excited mainly by starlight and they would not be produced by recombination of Si$^{+2}$. A similar situation occurs with the UV permitted lines of  \ion{O}{iii} $\lambda\lambda$ 3443,3759,3757 and others, which have been declared to be excited  by the Bowen mechanism \citep{bowen:35}, therefore they would not correspond to lines from recombination of O$^{+3}$. The efficiency of such a mechanism is not clear however \citep{kaler:67}, thus we have decided to keep these lines in our analysis  considering them as from pure recombination,  in order to determine the kinematical behaviour of the ions emitting them. 

Considering the velocities given by CELs (blue dots), in general it is found that most nebulae present a clear gradient in velocity, as a consequence of the ionisation structure in an expanding shell where the highly ionised species, which due to the ionisation structure are the nearest to the central star,  show the lowest expansion velocities while the low ionisation species show larger V$_{exp}$. That is, the expansion velocity increases with the distance to the central star in agreement with hydrodynamical models for PNe by e.g.,  \citet[][]{schonberner:99,schonberner:00}, which always show velocities increasing with the distance to the central star.  These models are computed by assuming that PNe are modelled by the interaction of the stellar winds of a unique central star. Thus the kinetic energy in the expanding shell can be seen as the original kinetic energy of the AGB wind, plus the amount deposited by the fast wind from the post-AGB PN core. The original AGB wind has a near-constant velocity with radius; and in hydrodynamical models the gradient in the PN velocity field is caused by the interacting winds and the ionisation structure  \citep{schonberner:14}. 

 The PNe showing this increase in V$_{exp}$(CELs)  with radius are   He\,2-86, M\,1-30, Pe\,1-1, 
Hb\,4,  M\,1-61, NGC\,3918, NGC\,7009,  PC\,14,  and NGC\,2867,   that is, 9 objects of a total of 14. Also Cn\,1-5 and M\,1-25 show  increasing V$_{exp}$ with the radius, although in these cases it is marginal, because only the low ionisation species [\ion{Fe}{ii}], [\ion{Fe}{iii}] and    [\ion{Ni}{ii}] are showing high expansion velocities, all the others (including [\ion{S}{ii}], [\ion{Cl}{ii}], [\ion{N}{ii}], and  [\ion{O}{ii}]) show V$_{exp}$ values similar to the ones of highly ionised species.

In our sample there are some objects not showing a gradient in the velocities given by  CELs. They are M\,3-15, M\,1-32, and PB\,8. These cases will be discussed in detail later, but here we say in advance that M\,3-15 and M\,1-32 show a very low V$_{exp}$ of about 11 -- 14 km s$^{-1}$ and  this occurs because both objects present a toroidal form where the toroid is on or near the plane of the sky \citep{rechy:17}, therefore the expansion velocity of the plasma in the toroid  is almost perpendicular to the line of sight.
Also PB\,8 shows a very flat gradient, with V$_{exp}$ of about 15 km s$^{-1}$, except for the highest ionizing species that have ver low V$_{exp}$. Probably the flat gradient is due to this compact nebula has an ellipsoidal morphology and it could be a relatively flat system oriented on or  near the sky plane.

A very interesting result, which is evident from an inspection of Figs. \ref{fig:graphics-1} and \ref{fig:graphics-2}, is that in most cases the expansion velocities measured from ORLs (red points)  appear smaller than the values given by CELs for the same ionisation potential. Besides in several cases V$_{exp}$(ORLs) show no gradient or an almost  flat gradient and such a gradient runs below or crosses the gradient observed for CELs. Therefore, for these cases, the velocity field derived from CELs is incompatible with the velocity field given by ORLs. This is evident if ORLs and CELs velocities for the same ion (for instance O$^{+2}$) are compared (see columns 6 and 7 in Table \ref{tab:velocities}). In most cases both velocities do not coincide as expected, and V$_{exp}$(CELs) are larger than V$_{exp}$(ORLs).  This would be indicating that, for a certain ion, its ORLs would be emitted mainly in zones closer to the central star than its CELs. 

It should be mentioned here that V$_{exp}$(ORLs)  of \ion{Si}{ii} and \ion{O}{iii} (whose permitted lines were considered to be produced by  recombination) show no particular discrepancy
in position relative to other ions with similar IP. On the contrary, if we assume that the \ion{Si}{ii} lines are excited by starlight  their position in the V$_{exp}$ vs. IP  figures, should be moved to IP=8.15 eV (IP value for Si$^+$) and this would produce a large discrepancy with the V$_{exp}$ of ions emitting CELs at the same IP. The same occurs for \ion{O}{iii}; if its lines are excited by the Bowen mechanism, its position in the V$_{exp}$ vs. IP diagram should be moved to IP=35.12 eV (IP value for O$^{+2}$) and in several cases  this would produce a large discrepancy with the points corresponding to V$_{exp}$[\ion{O}{iii}] and V$_{exp}$(\ion{O}{ii}) which are at this IP. We discuss this for each object individually.
\smallskip

Objects where  both, V$_{exp}$(CELs) and V$_{exp}$(ORLs), show a gradient  are: He\,2-86,  M\,1-30, Pe\,1-1, Hb\,4,  and NGC\,2867,   and most of the times V$_{exp}$(ORLs) are below V$_{exp}$(CELs) at the same IP.  The implications of this will be discussed in detail in next sections.

 Objects where V$_{exp}$(ORLs) show no gradient or a very flat one are: Cn\,1-5 where V$_{exp}$(ORLs) are in the range 24 $-$ 26 km s$^{-1}$, M\,1-25 with V$_{exp}$(ORLs)  of about 19 km s$^{-1}$,  M\,1-61 with V$_{exp}$(ORLs) $\leq$ 16 km s$^{-1}$, and NGC\,7009 with V$_{exp}$(ORLs)  $\sim$ 17 -- 18 km s$^{-1}$ (without considering \ion{He}{i} which is an ion distributed in a large zone of the nebula). Thus, in these objects  the ions emitting ORLs present about the same expansion velocity, independently  of their ionisation degree.  This is a very unusual situation in a PN plasma,  indicating that the ionised gas emitting ORLs is in a zone with a constant expansion velocity. 
This gas is rich in heavy elements and it is already ionised but possible it is not in ionisation equilibrium, therefore  it does not show the usual stratification found in  photoionised plasmas.  
This gas is  expanding at a lower velocity than the plasma emitting CELs, therefore it is possible that this gas had been ejected by the central star in a subsequent epoch  than the ejection of the low metallicity gas.

Some interesting objects in our sample are those showing the lowest ORL  velocities. They are: M\,1-61 which presents  V$_{exp}$(ORLs) $\sim$ 15 km s$^{-1}$ while  V$_{exp}$(CELs) go from 10 to 22 km s$^{-1}$, this nebula is ionised by a {\it wels},  it has a density n$_e$= 22200 cm$^{-3}$ and a radius of 0.029 pc; He\,2-86 which presents V$_{exp}$(ORLs) from 7 to 9 km s$^{-1}$  and V$_{exp}$(CELs) from 8 to 14 km s$^{-1}$, it is ionised by a [WC\,4] star, has a density n$_e$=23300 cm$^{-3}$ and a radius of 0.035 pc;   and Pe\,1-1, which has a density of n$_e$=31100 cm$^{-3}$, a radius of 0.039 pc, and presents also a small velocity gradient for ORLs, from 11 to 14 km s$^{-1}$  while V$_{exp}$(CELs) go from 10 to 17 km s$^{-1}$. These three PNe are the most compact ones in our sample, with a radius smaller than 0.04 pc, and they are very dense, therefore they should be  the youngest PNe analised here.  Other object with high density is M\,1-25. This PN  shows a very flat gradient for ORLs (from 18.6 km s$^{-1}$ for \ion{Si}{ii}  up to 20.2 km s$^{-1}$ for  \ion{O}{ii}), and it has a density of  n$_e$=15100 cm$^{-3}$, a radius of 0.052 pc and it is ionised by a [WC\,5-6] star. All these objects are compact, young and dense and their V$_{exp}$ are very low, in particular the ones calculated from ORLs. Other similarity among these PNe is that the difference (V$_{exp}$[\ion{O}{iii}] $-$ V$_{exp}$(\ion{O}{ii})) is smaller than 0.4 km s$^{-1}$. In M\,1-25 this difference is negative ($-$0.9 km s$^{-1}$). Therefore, it is found that the youngest PNe in our sample show the lowest ORLs expansion velocities.

From all the above we conclude that our results indicate, in many cases, that ORLs and CELs present dissimilar  kinematics and ORLs and CELs of the same ion seem to be produced in different zones of the nebula. ORLs, showing lower V$_{exp}$, would be emitted mainly in zones closer to the central star than the gas emitting CELs. 

\section{The {\it FWHM} of lines}

In the cases where two components,  approaching and receding, were found, the measurements were made separately by fitting two Gaussian profiles, thus  the line width of each component is not affected by the expansion velocity. Only turbulence, thermal broadening and nebular structure along the line of sight are included in the {\it FWHM}. 

In most cases occur that the {\it FWHM} of ORLs is narrower than the {\it FWHM}  of CELs. This is evident by comparing the {\it FWHM} of ORLs and CELs coming from the same ion, for instancerecombination  lines of \ion{O}{ii} compared with forbidden lines of [\ion{O}{iii}],  or lines of \ion{Ne}{ii} and [\ion{Ne}{iii}]. Velocities and {\it FWHM} of lines of \ion{O}{ii} and [\ion{O}{iii}] are presented in Table \ref{tab:velocities}. For the objects with split components, we have included only the {\it FWHM} of the blue component, because the values and behaviour of the red component are very similar.

If a similar turbulence is assumed for all the plasma, the lower {\it FWHM}  of ORLs is indicating that  the material emitting these lines is at a lower temperature than the gas emitting CELs. Our resut is in agreement with values of temperatures measured for ORLs and CELs where T$_e$(ORLs) is in general lower, on occasions  by several thousand degrees, than the temperature derived from CELs \citep[see e.g.,][]{wesson:05, zhang:05, mcnabb:13, fangliu:13}. In several PNe T$_e$ has been estimated from lines of \ion{He}{i} and from recombination lines of O$^{+2}$ and N$^{+2}$. These T$_e$(ORLs) are in average, several hundred degree lower than T$_e$(CELs).  The gas emitting ORLs is not necessarily  a H-deficient gas, but it should be a gas richer in heavy elements than the one emitting CELs, in order to be at lower temperature.

\begin{table*}
\caption{ [\ion{O}{iii}] and \ion{O}{ii} expansion velocities and  {\it FWHM} of the lines in the studied PNe \label{tab:velocities}}
\begin{tabular}{lccrrrrcccccc}
\hline 
\hline \\
PN G & name & profile &(V$_{hel})^c$~~ & V$_{exp}$(HI) & V$_{exp}$[\ion{O}{iii}]&V$_{exp}$(\ion{O}{ii}) &  FWHM& FWHM\\ 
  &  &  & km s$^{-1}$ &    km s$^{-1}$ & km s$^{-1}$ & km s$^{-1}$ & ([\ion{O}{iii}])\AA &(\ion{O}{ii}) \AA \\
\hline\\
002.2$-$09.4 & Cn\,1-5$^b$ & split & $-1.8\pm$2.4 &25.0$\pm$0.2& 26.4$\pm$1.2 & 25.0$\pm$0.7 &0.47$\pm$0.02& 0.35$\pm$0.09\\
003.1+02.9& Hb\,4$^b$ & split &$-66.2\pm$9.5& 17.2$\pm$1.6&16.1$\pm$1.0&11.2$\pm$0.2&0.43$\pm$0.01&0.30$\pm$0.04\\
004.9+04.9 & M\,1-25$^b$ &split& 9.2$\pm$2.6 &19.1$\pm$0.3 & 19.2$\pm$0.3 & 20.2$\pm$0.3 &0.53$\pm$0.02 & 0.44$\pm$0.08\\
006.8+04.1 & M\,3-15$^b$ &split & 156.0$\pm$4.9 &12.4$\pm$3.0 & 11.9$\pm$0.3 &14.2$\pm$1.0 & 0.41$\pm$0.03 & 0.45$\pm$...\\
011.9+04.2 & M\,1-32$^a$&single& $-90.9\pm$3.6 & 17.1$\pm$0.1 & 12.5$\pm$0.7 & 8.4$\pm$1.0 & 0.40$\pm$0.04 & 0.26$\pm$...\\ 
019.4$-$05.3& M\,1-61$^a$ & single & 5.61$\pm$3.3& 18.5$\pm$0.4 & 15.0$\pm$0.3 & 14.7$\pm$0.9 &0.48$\pm$0.05 & 0.43$\pm$0.02\\ 
037.7$-$34.5& NGC\,7009 & split & --- & 19.6$\pm$0.1 & 19.1$\pm$0.7 & 18.0$\pm$0.4 & --- & ---\\
278.1$-$05.9& NGC\,2867$^b$& split & 14.0$\pm$3.1 &28.1$\pm$0.3&27.3$\pm$0.3 &24.5$\pm$0.1& 0.25$\pm$0.02& 0.27$\pm$0.02 \\
285.4+01.5 & Pe\,1-1$^b$  &split  & 47.3$\pm$3.4 & 11.0$\pm$0.1 & 11.2$\pm$0.1 & 10.9$\pm$0.1 &0.33$\pm$0.03 & 0.28$\pm$0.02\\
 292.4+04.1 & PB\,8$^a$ & single & 16.8$\pm$4.7 &17.0$\pm$0.2 & 14.5$\pm$0.2& 12.8$\pm$1.1 &0.46$\pm$0.05 &0.38$\pm$0.05\\
294.6+04.7 & NGC\,3918 &split  &$-19.1\pm$2.8  &14.1$\pm$0.5 & 13.9$\pm$0.4 & 15.2$\pm$0.1 & 0.30$\pm$0.02 & 0.30$\pm$0.01\\
300.7$-$02.0 & He\,2-86$^b$ & split& 4.6$\pm$3.6 & 9.1$\pm$1.5 & 7.9 $\pm$0.3&7.5$\pm$0.9 & 0.19$\pm$0.05 & 0.23$\pm$0.05\\
336.2$-$06.9& PC\,14$^b$& split & $-33.5\pm$8.4 &21.7$\pm$0.1 &22.2$\pm$0.4 & 20.9$\pm$1.9 &0.46$\pm$0.07 &0.39$\pm$0.09\\
355.9$-$04.2 & M\,1-30$^b$ & split & $-125.5\pm$3.1 & 11.6$\pm$0.2 &10.3$\pm$0.1 & 9.2$\pm$0.5&0.32$\pm$0.03 & 0.22$\pm$0.05\\
\hline 
\multicolumn{7}{l}{$^a$ For  single lines, V$_{exp}$ is derived from 1/2 {\it FWHM}.}\\
\multicolumn{7}{l}{$^b$ For split lines, {\it FWHM} from the blue component is listed.}\\
\multicolumn{9}{l}{$^c$ Heliocentric velocities, V$_{hel}$, have been calculated as the average of V$_{rad}$ of all the measured lines,}\\
\multicolumn{9}{l}{\hskip 0.2cm corrected for Earth movement.}
\end{tabular}
\end{table*}

\section{Individual objects}

\subsection{PN G002.2-09.4  Cn\,1-5}
This PN, ionised by a [WO\,4]pec star, is N-rich (a Peimbert's Type I PN) and  shows an ADF(O$^{+2}$) of 1.9. Its radius of 0.079 pc and age of 4300 yr indicate a relatively evolved nebula. \citet{sahai:11}  classified it as ``B,o bcr(o,i)'' pointing out that it has a bipolar structure with open lobes, and at the centre  it shows a bright barrel shape with open ends and an irregular structure in the interior. 

 The slit crossed the central position and all the lines (CELs and ORLs) are well split (see Fig. \ref{fig:profiles}), therefore we fit a Gaussian profile to the blue and red components independently.

Partial results for this object have been published in the Proceedings  of IAU Symposium No. 323 (Pe\~na et al. 2017).

In this work V$_{exp}$ of the ions as a function of IP  are shown in Fig. \ref{fig:graphics-1}.  CELs show a mild gradient in velocity.  The velocities go from V$_{exp}$ $\sim$ 25 $-$ 26 km s$^{-1}$ near the central star  (highly ionised species) to  V$_{exp}$ of about 30 $-$ 32 km s$^{-1}$ for the low ionised species ([\ion{Fe}{ii}], [\ion{Fe}{iii}] and [\ion{Ni}{ii}]).

On the other hand, expansion velocities from ORLs show a flat gradient. Lines from  \ion{Si}{ii} (considered as due to  recombination) present  V$_{exp}$ of about 25 km s$^{-1}$ and similar values are found from  \ion{N}{ii}, \ion{O}{ii} and \ion{Ne}{ii} lines.  In general V$_{exp}$ of ORLs are below V$_{exp}$ of CELs at the same IP.  In particular V$_{exp}$[\ion{O}{iii}] and V$_{exp}$[\ion{Ne}{iii}] are larger than V$_{exp}$(\ion{O}{ii}) and V$_{exp}$(\ion{Ne}{ii}) respectively.

Line widths are larger for CELs than for ORLs, for the blue component {\it FWHM}([\ion{O}{iii}])=0.47$\pm$0.02 \AA,  while {\it FWHM}(\ion{O}{ii})=0.35$\pm$0.09 \AA. The only discrepant lines correspond to the permitted lines of \ion{C}{ii}, which show a  large V$_{exp}$ of 30 km s$^{-1}$ and also present a large
{\it FWHM} of 0.60 \AA. 

\subsection{PN G003.1+02.9 Hb\,4}

This high-excitation compact PN shows a high N/O abundance ratio of 0.70, that classifies it as a Peimbert's Type I PN. With a physical radius of 0.06 pc and an age of about 3000 yr it is a relatively  young  PN. \citet{garcia:13}  reported an ADF(O$^{+2}$)=3.7, one of the largest in the studied sample. 

{\it HST} images in [\ion{N}{ii}] and H$\alpha$ lines show a very filamentary ring surrounded by a faint multipolar  halo. Two elongated ionised  knots (ansae or collimated outflows) at high velocity ($\pm$150 km s$^{-1}$)  are found at both sides of the nebula \citep{lopez:97}.
\citet{sahai:11} classify it as `M,c bcr(i)" (multipolar, close borders, barrel shape central region, irregular inside). Also it shows  point-symmetric features and ansae.  

Our spectra cover the central part of the nebula, passing though the central star position. The lines have complex profiles and appear slightly split therefore they can be deblended showing expansion velocities  similar or smaller than 20 km s$^{-1}$ in the case of CELs and even smaller in the case of ORLs.
 The recombination line \ion{O}{ii} $\lambda$4649 has 2 close components, similar to the CEL [\ion{O}{iii}] $\lambda$4363, but with a lower V$_{exp}$ of 12.5 km s$^{-1}$ versus 15.9 km s$^{-1}$ for [\ion{O}{iii}].
 
In  the V$_{exp}$ vs. IP graph (Fig. \ref{fig:graphics-2}),  a gradient due to the velocity structure within the nebula is very clear for CELs. V$_{exp}$ is about 10 km s$^{-1}$ for  the highly ionised species like Mn$^{+4} $ and Ar$^{+4}$ and it increases up to about 20 km s$^{-1}$ for O$^+$, N$^+$, S$^+$, Cl$^+$, and other ions located far from the star.

V$_{exp}$(ORLs) vs. IP graph also shows a gradient but flatter  than the case of CELs.  At any ionisation potential, V$_{exp}$(CELs) are larger than  V$_{exp}$(ORLs) except for the case of \ion{He}{ii}. In particular V$_{exp}$(ORLs) from low ionised species like \ion{Si}{ii} and \ion{C}{ii} are much lower than velocities of CELs at the same IP. Notice that if the \ion{Si}{ii} lines are considered as excited by starlight instead of recombination, this point should be moved to IP=8.15 eV, making the discrepancy even larger. 

For  O$^{+2}$, it is found that V$_{exp}$[\ion{O}{iii}]=16.1$\pm$1.0 km s$^{-1}$ while V$_{exp}$(\ion{O}{ii})=11.2$\pm$0.2 km s$^{-1}$. This nebula also shows permitted lines of \ion{O}{iii}, in the graph they are considered as from recombination  of O$^{+3}$ and have an average V$_{exp}$ of 7.6 km s$^{-1}$. If they were excited by the Bowen mechanism, the point should be moved to IP=35.12 eV and it would be very discrepant with both, V$_{exp}$[\ion{O}{iii}] and V$_{exp}$(\ion{O}{ii}). Therefore the permitted lines of \ion{O}{iii} seem to be produced mainly by recombination.

The kinematical behaviour  of CELs and ORLs indicates that ions emitting ORLs would be closer to the central star than those emitting CELs. As in Cn\,1-5, line widths are larger for CELs than for ORLs, for the blue component.

\subsection{PN G004.9+4.9  M\,1-25}
This is a young, compact  and dense PN ionised by a [WC\,5-6] central star. It has a very moderate ADF(O$^{+2}$)  of 1.51 and its abundances are normal (N/O=0.3). 
The {\it HST} image shows a bright irregularly elongated ring surrounded by a faint halo. \citet{sahai:11} classified it  as ``E,c  h(e,d)" (elongated, closed lobes with an  elongated halo, with a sharp outer edge).

Its V$_{exp}$ vs. IP graph (Fig. \ref{fig:graphics-1}) shows a  gradient for the CELs  mainly due to the  [\ion{Ni}{ii}], [\ion{Fe}{ii}] and [\ion{Fe}{iii}] lines are showing V$_{exp}$ of 26 --  24 km s$^{-1}$. All the other ions show smaller CEL velocities between 19 and 20 km s$^{-1}$.  On the other side, V$_{exp}$(ORLs) graph presents a very flat behaviour, with velocities between 18 to 20 km s$^{-1}$.  The lines from \ion{Si}{ii} show a V$_{exp}$ lower than CELs at the same IP.  Moving the \ion{Si}{ii} point to a lower IP of 8.15 eV would increase the discrepancy.

The expansion velocities  as given by CELs and ORLs do not coincide, especially in the outer zone. In this case, V$_{exp}$[\ion{O}{iii}] is slightly lower than V$_{exp}$(\ion{O}{ii}).

Line widths are larger for CELs than for ORLs, for the blue component.

\subsection{PN G006.8+04.1  M\,3-15 and PN G011.9+04.2  M\,1-32}

These objects are discussed together because they present very similar morphology  and also present some similar kinematical characteristics, as discussed by \citet{rechy:17}. Both objects belong to the galactic bulge.

In both cases the PNe are constituted by a toroid almost contained in the plane of the sky (pole-on toroid) with ejections escaping from the poles at high velocities. The jets get a velocity of about $\pm$180 km s$^{-1}$ for M\,1-32 and of about $\pm$90 km s$^{-1}$ for M\,3-15. \citet{sahai:11} describe M\,3-15 as ``L,c bcr(c)", thus it has a pair of collimated lobes, and a barrel shape structure in the centre which would conform the toroid. 

The fact that the main body of the nebulae, the toroid, is contained in the plane of the sky makes them to present low expansion velocities and the gradient of the velocity field is not noticeable.  However there are interesting differences regarding the ORLs and CELs behaviour.

For M\,3-15 the graph V$_{exp}$ vs. IP  (Fig. \ref{fig:graphics-1}) is very flat for CELs and for ORLs. V$_{exp}$ have a value of about  11 $-$ 13 km s$^{-1}$ for all the ions emitting CELs.  On the other hand ORLs show also a flat but disperse behaviour with V$_{exp}$ for \ion{Si}{II}, \ion{C}{II}, \ion{He}{I}, and \ion{Ne}{ii} between 12 and 13 km s$^{-1}$, while \ion{O}{ii} and \ion{N}{iii} present V$_{exp}$ of 14.2 and 15.9 km s$^{-1}$  respectively. In this nebula ORLs present higher V$_{exp}$ than CELs, which is very noticeably in the cases of [\ion{O}{iii}] compared to \ion{O}{ii} and [\ion{Ne}{iii}] compared to \ion{Ne}{ii}.

The profiles in M\,1-32 present a very intense single component coming from the toroid and faint extended high-velocity wings  (see Fig.\ref{fig:profiles} at the bottom).  For this work we have measured only the bright single component.   V$_{exp}$ vs. IP graph for CELs shows a similar behaviour to the one of M\,3-15. There is no gradient in velocity and all the ions present V$_{exp}$ between 11 and 14 km s$^{-1}$, except [\ion{Ca}{ii}] with V$_{exp}$= 16.6 km s$^{-1}$. ORLs show also a flat V$_{exp}$ gradient with values around 11 to 14 km s$^{-1}$ except V$_{exp}$(\ion{O}{ii}) which is 8.4 km s$^{-1}$.  In this nebula V$_{exp}$[\ion{O}{iii}] (12.5 km s$^{-1}$) and V$_{exp}$(\ion{O}{ii})  are very discrepant.     

 Thus, in both nebulae the kinematics of CELs emitted by the toroidal component are similar, but the ORLs present a different kinematics than CELs and dissimilar in both cases.

The {\it FWHM} for CELs is larger than for ORLs in M\,1.32, but not in M\,3-15.

The analysis of these objects for which  hydrodynamical models have been computed allows us to understand  the morphology and kinematics of other PNe in this sample, whose V$_{exp}$ vs. IP graphs for CELs present a similar behaviour.

\subsection{PN G019.4-05.3  M\,1-61}
This very compact, young and dense PN shows a very complex morphology classified as ``M,c bcr,  ps, ml, ib, h" (multipolar with close lobes, a barrel shape inside, point symmetric, minor lobes, inner bubble, and halo) by \citet{sahai:11}. It presents a moderate ADF(O$^{+2}$) of 1.66 and its abundance ratio 
N/O=0.40 is normal. The central star is a weak emission line star ({\it {\it wels}}).

V$_{exp}$ given by CELs show a clear gradient, with values from about 12 km s$^{-1}$ in the zone near to the star up to 24 km s$^{-1}$ in the outer zone. Velocities measured from ORLs show a flatter  gradient with  values from  14 to 16 km s$^{-1}$. Both gradients come across at the level of  O$^{+2}$ ionisation potential, where the velocities of CELs and ORLs coincide. 

In the low ionisation extreme V$_{exp}$ from CELs (e.g., [\ion{N}{ii}] and [\ion{O}{ii}])  are well above  V$_{exp}$(\ion{Si}{ii}).  If the lines of \ion{Si}{ii} were excited by stellar photons,  the \ion{Si}{ii} point should be moved to a lower IP of 8.15 eV, increasing by much the difference of this point with  the ones of [\ion{N}{ii}] and [\ion{O}{ii}].

This object shows the permitted lines of \ion{O}{iii}. If they are considered as from recombination of O$^{+3}$, its V$_{exp}$ position agrees with the gradient showed by ORLs. If they are considered as excited by the Bowen mechanism, the point should be moved to IP=35.12 eV where it would coincide, within uncertainties, with the points representing [\ion{O}{iii}] and \ion{O}{ii}, thus in this case we can not affirm if these lines are from recombination or excited by the Bowen mechanism.

Again in this case, the velocity fields from ORLs and CELs are incompatible. ORLs seem to be emitted in a different plasma, possibly closer to the central star.

\subsection{PN G037.7-34.5 NGC\,7009}
This is an extended and complex multishell nebula (radius of 0.076 pc), ionised by a normal hot  central star. Its age is about 3700 yr.  \citet{liu:95}  determined an ADF(O$^{+2}$) of about 5 (the largest in this PN sample). Spatially and velocity-resolved echelle spectroscopy was analysed by \citet{richer:13} who found that the kinematics of recombination lines (\ion{C}{ii}, \ion{N}{ii}, \ion{O}{ii}, and \ion{Ne}{ii} lines) does not coincide with the  kinematics of CELs ([\ion{O}{iii}] and [\ion{Ne}{iii}] lines). These authors propose that in this nebula there is an additional plasma component and that the recombination  lines arise in a different volume from that giving rise to the forbidden emissions from the parent ions.  

By using \citet{richer:13} data we measured  the splitting of lines and plot the V$_{exp}$ vs. IP graph  in Fig. \ref{fig:graphics-2}. This graph shows that CELs have slightly larger expansion velocities than ORLs of the same ion (e.g., see the cases of [\ion{O}{iii}] and \ion{O}{ii}, [\ion{Ne}{iii}] and \ion{Ne}{ii}).  That is, the kinematics of CELs and ORLs differs in the same sense that in the other objects. Velocity fields of ORLs and CELs are incompatible confirming the results by \citet{richer:13}.

This nebula also shows permitted lines of \ion{O}{iii}. In the graph they are considered as produce by  recombination and show an average velocity of 17.4$\pm$0.3  km s$^{-1}$. If they were excited by the Bowen mechanism, the point should be moved to IP=35.12 eV and it would be  discrepant with  [\ion{O}{iii}] (V$_{exp}$=19.1$\pm$0.7 km s$^{-1}$) , although it could be considered similar to \ion{O}{ii} (V$_{exp}$=18.0$\pm$0.4 km s$^{-1}$), within uncertainties. Therefore the permitted lines of \ion{O}{iii} could be produced by the Bowen mechanism when compared to \ion{O}{ii} recombination lines.

\subsection{PN G278.1-05.9 NGC\,2867}

This is an extended nebula with a projected diameter of  14$''$ and a physical diameter of 0.16 pc. It is ionised by a hot [WC\,2] star, therefore the nebula shows high excitation including [\ion{Ne}{v}] lines. Its morphology is  ``E sm''  \citep[elliptical with structures and multiple shells,][] {gorny:95} and its age seems to be no larger than 2750 yr.
 \citet{garcia:09}  derived a moderate ADF(O$^{+2}$) of about 1.58. for this nebula.
 
 The slit crossed through the central star, but the spectrum was extracted from a bright knot near the  star \citep{garcia:09}. 
The nebula is well resolved, and it presents well split lines. Red and blue components were measured independently.

Fig. \ref{fig:graphics-2} hows V$_{exp}$ measured for different ions vs. the IP. Velocities from  CELs present a well defined gradient, V$_{exp}$ go from 20 km s$^{-1}$ in the inner zone (ion Ne$^{4+}$) up to 33 km s$^{-1}$ in the outer zone (ions O$^+$, N$^+$, S$^+$). Velocities from ORLs also show a gradient but  V$_{exp}$(CELs) are  in general larger than V$_{exp}$(ORLs) at the same ionisation potential, specially in the low ionisation zone.
This is particularly true for CELs from [\ion{O}{iii}] and ORLs from \ion{O}{ii} which present V$_{exp}$ of 27.3$\pm$0.3 km s$^{-1}$ and 24.2$\pm$0.1 km s$^{-1}$ respectively. In this case, the {\it FWHM}(CELs) and {\it FWHM}(ORLs) are similar (0.25 \AA~ and 0.27 \AA~ respectively).

This nebula shows permitted lines of \ion{O}{iii}. They are considered as produced by recombination of O$^{+3}$ and present an average velocity of 23.8$\pm$0.9 km s$^{-1}$. If they were excited by the Bowen mechanism, the point should be moved to IP=35.12 eV and it would be discrepant with [\ion{O}{iii}] (V$_{exp}$= 27.3$\pm$ 0.3) but it would coincide within uncertainties with  \ion{O}{ii} (V$_{exp}$=24.2$\pm$0.1). Therefore the permitted lines of \ion{O}{iii} could be excited by the Bowen mechanism.

\subsection{PN G285.4+01.5 Pe\,1-1}
This is a very compact, dense and young PN ionised by a [WC\,4] central star. Its ADF(O$^{+2}$) is 1.70 and its N/O abundance ratio of 0.25 is normal. It was classified by \citet{sahai:11}
as bipolar, with close lobes, and a barrel shape structure in the center. Our spectrum comes from the bright central structure and the slit passed through the central star. The barrel shape is noticeable in the sense that CELs and ORLs appear slightly split  and can be safely deblended.

The expansion velocity is low (no larger than 17 km s$^{-1}$ for the low ionisation species), and a clear gradient in the velocity field, as given by CELs, is detected. A similar gradient is shown by ORLs. Expansion velocities almost coincide for CELs and ORLs, although in general V$_{exp}$(ORLs) are slightly lower than V$_{exp}$(CELs). 

 The permitted lines of \ion{Si}{ii} are considered as produced by recombination and   V$_{exp}$(\ion{Si}{ii})  is well located in comparison with V$_{exp}$ of [\ion{N}{ii}] and [\ion{O}{ii}]. If the lines were excited by starlight  the \ion{Si}{ii} point should be  moved to IP=8.15 eV and it would be discrepant with the CELs at this IP.

\subsection{PN G292.4+04.1 PB\,8}
This PN is ionised by a peculiar [WC]/[WN] central star \citep{todt:10}. It has a moderate ADF(O$^{+2}$) of 2.19 and a normal chemical composition with N/O = 0.27. Its radius is extended and its age is about 3500 yr. It shows single lines, and no gradient is detected in the velocity field. All the expansion velocities are lower than 16 km s$^{-1}$. In particular V$_{exp}$[\ion{O}{iii}] $\sim$ 14 km s$^{-1}$ and  V$_{exp}$(\ion{O}{ii}) $\sim$ 13 km s$^{-1}$. The highly ionised species (lines from [\ion{Ar}{iv}] and \ion{N}{iii}) show even lower velocities.   

This nebula was classified as an ``elliptical compact PN" \citep{schwarz:92}. There is no high-resolution image of this object but due to the behaviour of its velocity field, and considering the cases of M\,1-32 and M\,3-15,  it is probably a  pole-on toroid.

\subsection{PN G294.6+04.7  NGC\,3918}
This highly ionised PN has a complex point-symmetric morphology. It presents a moderate ADF of about 1.80 and its N/O abundance ratio of 0.28 is normal. The nebula was deeply studied by \citet[][G-R2015]{garcia:15} from a high-resolution spectrum (R $\sim$ 40,000) obtained with UVES at VLT, aiming to analyse the s-processes in the central star. The same data is employed here to analyse the kinematics of CELs and ORLs. The spectrum was obtained from a slit passing at 3.8  arcsec North  from the central star, oriented E-W  (see Fig. 1 by G-R2015). In this sense the expansion velocity measured is not the real expansion velocity, but the radial projection of the velocity at that point.

The lines show a well split profile, therefore we computed V$_{exp}$ from  CELs and ORLs by fitting a Gaussian profile to each one of the components. 

The graph V$_{exp}$ vs. IP shows a clear gradient in the velocity field as given by CELs. The lines from the most highly ionised species, [\ion{K}{iv}], [\ion{Ar}{iv}], [\ion{Cl}{iv}] and [\ion{Kr}{iv}], show small V$_{exp}$ from 9 to 11 km s$^{-1}$. At an intermediate V$_{exp}$ between 14 $-$18 km s$^{-1}$, we find [\ion{Ne}{iii}], [\ion{O}{iii}], [\ion{Ar}{iii}], [\ion{S}{iii}], [\ion{Cl}{iii}] and [\ion{Fe}{iii}]. The largest  velocities, from 22 to 23 km s$^{-1}$, occur for [\ion{N}{ii}], [\ion{O}{ii}], [\ion{Cl}{ii}] and [\ion{S}{ii}].  
On the other hand, V$_{exp}$(ORLs) show a flatter gradient where  ions with IP from 20 to 40 eV present  V$_{exp} \sim$ 15 km s$^{-1}$ while the most ionised species show V$_{exp}$ of about 12 km s$^{-1}$.  For this object V$_{exp}$[\ion{O}{iii}] is slightly lower than V$_{exp}$(\ion{O}{ii}).

This nebula  shows the permitted lines of \ion{Si}{ii}. If these lines are produced by recombination, V$_{exp}$(\ion{Si}{ii}) coincides well with V$_{exp}$ of [\ion{Fe}{iii}]. If the point for \ion{Si}{ii} is moved to 8.15 eV, it would be discrepant with V$_{exp}$ of [\ion{S}{ii}] and others.
Also V$_{exp}$ of permitted lines of \ion{O}{iii} are plotted in the graph,  they are considered as from recombination and have an average velocity of 11.4$\pm$0.3 km s$^{-1}$. If these lines were excited by the Bowen mechanism, the point should be moved to IP=35.12 eV and it would be very discrepant with both, [\ion{O}{iii}] (V$_{exp}$=13.9$\pm$0.4)  and \ion{O}{ii} (V$_{exp}$=15.2$\pm$0.1). Therefore the permitted lines of \ion{Si}{ii} and \ion{O}{iii} seem to be produced mainly by recombination of Si$^{+2}$ and O$^{+3}$ respectively.

\subsection{PN G300.7-02.0 He\,2-86}

This is a compact and young PN ionised by a [WC\,4] star. It is very dense and N-rich. It presents a moderate ADF(O$^{+2}$)\,=\,1.94. 
 \citet{sahai:11} classify it as ``M,c bcr(o)  ml, h(a)" indicating that the nebula is ``multipolar with close ends,  and a barrel shape with open ends, minor lobes, and halo".
In the HST image it looks bipolar with a toroid in the center.

CELs as well as ORLs appear almost blended. A deblend of two gaussians can be made, giving expansion velocities for CELs from about 7.8 km s$^{-1}$ for the highly ionised species to about 13 km s$^{-1}$ for the low ionised ones. Therefore there is a clear gradient in the velocity field presented by CELs (see Fig. \ref{fig:graphics-1}). The velocities given by ORLs are always slightly lower than the velocities of CELs at the same ionisation potential. Therefore, if we associate an expansion velocity with the distance to the central star,  again ORLs seem to be  emitted in a zone closer to the central star than CELs, for the same ions.

This nebula  shows the permitted lines of \ion{Si}{ii} coinciding well in expansion velocity  with CELs of [\ion{Fe}{iii}]. If the point is moved to 8.15 eV, it would be very discrepant with CELs of [\ion{S}{ii}], [\ion{Ni}{ii}] and others.

The low expansion velocities shown by the ions in this nebula seem to be related with the high density and youth of the nebula.

\subsection{PN G336.3-06.9 PC\,14}

This bipolar nebula is ionised by a [WO\,4] central star. It has an angular  diameter  of 7'', a heliocentric distance of 5796 pc, and a physical radius 0.087 pc \citep{frew:16}. Its age, derived from the R/v$_{exp}$ ratio is about 3900 yr.
The slit passed through the central star position.   The ADF(O$^{+2}$) is moderate with a value of  1.94 \citep{garcia:13}.

All the nebular lines (CELs and ORLs) are split in red and blue components which were measured independently.  
In Fig. \ref{fig:profiles}, the line profiles of [\ion{O}{iii}] $\lambda$4363 and \ion{O}{ii} $\lambda\lambda$4649,4650 are shown. It is evident that ORLs are narrower than the CELs (from Table \ref{tab:velocities}, {\it FWHM}([\ion{O}{III}]) = 0.46$\pm$0.07 \AA, while {\it  FWHM}(\ion{O}{II}) = 0.39$\pm$0.09\AA). 

Partial results for this object have been published in the Proceedings  of IAU Symposium No. 323 (Pe\~na et al. 2017). For completeness,  here we include  the graph V$_{exp}$ vs. IP (see Fig. \ref{fig:graphics-2}) where it is evident that V$_{exp}$(CELs)  follow the velocity field of an expanding shell, with velocities that go from about 20 km s$^{-1}$ in the inner zone (lines of [\ion{Ne}{iii}], [\ion{Ar}{iv}], [\ion{K}{iv}]), to 28 km s$^{-1}$ in the outer zone (lines of [\ion{Fe}{ii}], [\ion{S}{ii}], [\ion{N}{ii}]).  On the other hand ORLs show a gradient below the CELs gradient.  ORLs from intermediate ionised species (He$^+$, C$^{+2}$, N$^{+2}$, O$^{+2}$, Ne$^{+2}$, and C$^{+3}$) present a nearly constant velocity of about 20 $-$ 22 km s$^{-1}$ while ORLs from the highly ionised species O$^{+3}$, N$^{+2}$ and He$^{+2}$ have a low V$_{exp}$ of about 16 $-$ 18 km s$^{-1}$, producing the effect of a gradient. It seems evident that, as far as CELs and ORLs from the same ion do not share the same kinematics,  these lines are not being emitted in the same zone of the nebula. ORLs  seem to be emitted in an inner zone.

V$_{exp}$ from permitted lines of \ion{O}{iii} is plotted in the graph. These lines are considered as produced from recombination and have a V$_{exp}$  of 16.9$\pm$1.0 km s$^{-1}$. If they were excited by the Bowen mechanism, the point should be moved to IP=35.12 eV and it would be very discrepant with both, [\ion{O}{iii}] (V$_{exp}$=22.2$\pm$0.4)  and \ion{O}{ii} (V$_{exp}$=20.9$\pm$1.9). Therefore the permitted lines of \ion{O}{iii} seem to be produced mainly by recombination of O$^{+3}$.

\subsection{PN G 355.9-04.2, M1-30}

This is a very compact and young low-excitation PN  ionised by a weak emission line star ({\it wels}). \citet{sahai:11} classify it as ``M,c,t ps(m),h(a)" (multipolar with close lobes and a toroid, point-symmetric with opposite lobes, halo with arc-like structures).  %Age 1530 yr. 

\citet{garcia:13} determined an  ADF(O$^{+2}$) of 2.1 for this object.
The observed spectrum crosses through the central star  position. The heliocentric radial velocity is $-$125  km s$^{-1}$ indicating that this is an object of the galactic bulge. The chemical composition is normal with N/O abundance ratio of 0.49.

Lines appear split but very close. In general CELs and ORLs can be deblended, except \ion{C}{ii} $\lambda$4267  that is single, but with a complex profile.
The graph V$_{exp}$ vs. IP, presented in Fig.  \ref{fig:graphics-1}, shows clearly the velocity structure inside the nebula as given by CELs. The highly ionised species show expansion velocities of about  10  km s$^{-1}$ or lower, and the low ionised species show  V$_{exp}$ of about 17  km s$^{-1}$. Also V$_{exp}$ from ORLs show a gradient although, similarly to other cases, ORLs  present lower V$_{exp}$ than CELs at the same IP. In particular V$_{exp}$[\ion{O}{iii}] is 10.3$\pm$0.1 km s$^{-1}$ while V$_{exp}$(\ion{O}{ii}) is 9.2$\pm$0.5 km s$^{-1}$. On the other hand [\ion{Ne}{iii}] and \ion{Ne}{ii} have the same V$_{exp}$ within uncertainties. 

Given that V$_{exp}$ from ORLs are lower, it would be indicating that the ions emitting ORLs  lie inside the zone where CELs are produced.

This nebula  shows the velocity of permitted lines of \ion{Si}{ii} coinciding well with V$_{exp}$ of [\ion{Fe}{iii}]. If the \ion{Si}{ii} lines are excited by the starlight, the point should be moved to 8.15 eV and  it would appear much below V$_{exp}$ of [\ion{S}{ii}]  and others.
Therefore the permitted lines of \ion{Si}{ii} seem to be produced mainly by recombination of Si$^{2}$.

\section{General conclusions}

We have analysed the kinematics of ions emitting CELs and ORLs in fourteen PNe, ten of which are ionised by a [WC] central star, two are ionised by a {\it wels}, and two have normal central stars. 

High resolution spectra obtained mostly with the MIKE spectrograph attached to the Magellan telescope Clay  have been used, except for two objects whose data were obtained from the literature. Expansion velocities were determine from CELs and ORLs as half the difference in velocity between the red and blue components if the lines were split or as half the {\it FWHM} if the lines were single. V$_{exp}$ were analysed as a function of the distance of ions with respect to the central star, represented by the ionisation potential of the ion.

In most of the analysed objects, the emission from CELs shows a gradient in velocity as a function of the distance to the central star, which is expected  in an expanding  plasma with a structure determined  by  ionisation equilibrium and  the effect of interacting winds which shape the hydrodynamic structures.
We recall here that in hydrodynamical models of such a plasma  the velocity field accelerates outwards, the shell's edge expands faster than the inner zone, and both  accelerate as the star evolves with time crossing the HR diagram. See Fig. 4 by \citet{schonberner:16}.

Two objects which do not show such a gradient and present a CEL velocity field almost flat with a very low expansion velocity are M\,1-32 and M\,3-15 whose main emission comes from a toroid oriented almost pole-on, thus it is contained in the plane of sky. Therefore only a small  radial velocity  is measured for all the ions  in both cases.  Hydrodynamical models for H$\alpha$ and [\ion{N}{ii}] lines for both objects are very similar. In these objects ORLs do not show a velocity gradient either, but they present a different behaviour in the sense that, in M\,3-15 V$_{exp}$(ORLs) are larger than V$_{exp}$(CELs) and it is the opposite in M\,1-32. Therefore, in these two cases CELs and ORLs do not share the same kinematics.

Considering the expansion velocities derived from ORLs, we found that many of the analysed objects show a velocity gradient flatter than CELs or a gradient below that of CELs. Therefore the kinematics shown by ORLs is incompatible with that of CELs, specially if the emission from the same ions is considered (for instance the CEL and ORL emissions of O$^{+2}$ and Ne$^{+2}$).  
To illustrate these differences we present Fig.  4%\ref{fig:Vexp}
 where the expansion velocities from CELs emitted by O$^{+2}$ ([\ion{O}{iii}] Expansion velocity) are plotted against  V$_{exp}$ from ORLs emitted by the same ion (\ion{O}{ii} Expansion velocity). In this figure  it is evident  that in seven cases (represented in blue) V$_{exp}$[\ion{O}{iii}] is definitely larger than V$_{exp}$(\ion{O}{ii}). 
For the four cases marked in green  the uncertainty bars prevent us from a conclusive result, in these cases both V$_{exp}$ would be equal within uncertainties. Only three objects (NGC\,3918, M\,3-15, and M\,1-25 marked in red) show  V$_{exp}$[\ion{O}{iii}] slightly lower than V$_{exp}$(\ion{O}{ii}). In any case the important conclusion here is that  in most objects (10 of 14) both  velocities do not coincide and in the majority of cases V$_{exp}$(CELs) are larger than V$_{exp}$(ORLs).

\begin{figure} \label{fig:Vexp}
\begin{center}
\includegraphics[scale=0.55]{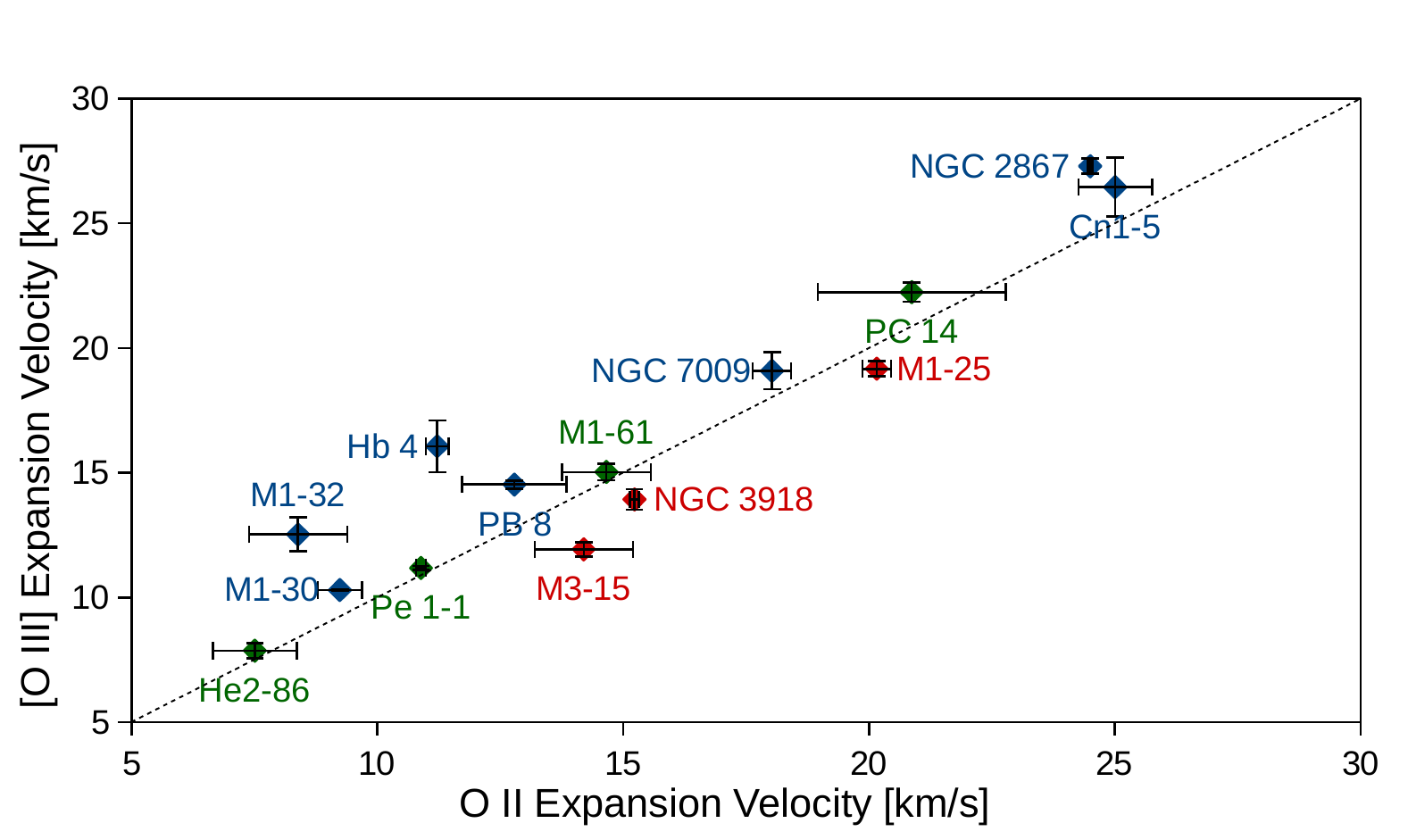}
\end{center}
\caption{V$_{exp}$ from [\ion{O}{iii}] lines are compared with V$_{exp}$ from \ion{O}{ii} lines. In seven  objects (in blue) V$_{exp}$[\ion{O}{iii}] is larger than V$_{exp}$(\ion{O}{ii}). In three (in red) it is the opposite and in four objects, both velocities are equal within uncertainties.}
\end{figure}

When a clear velocity gradient is found from CELs and a flatter  gradient (lying below the gradient from CELs) is observed from  ORLs,  this is indicating  that  the ions producing ORLs are located  in a zone more centrally concentrated that the zone where collisionally excited lines are emitted.
In the  cases where V$_{exp}$ from ORLs show  no gradient (no velocity structure), the ions emitting this radiation are behaving  differently to what is  predicted by hydrodynamical modelling  for a plasma  in ionisation equilibrium. 
The zone emitting ORLs  could correspond to gas ejection from the star at a later time than the ejection of the zone emitting CELs; this gas was ejected at a low velocity  and no acceleration due to interacting winds has occurred yet. 

We have analysed the kinematics shown by the permitted lines of \ion{Si}{ii} and \ion{O}{iii} which have been indicated as lines excited by starlight in the first case and Bowen mechanism in the second one. Our analysis reveals that for most of the studied objects  such lines seem to be produced  mainly by recombination of Si$^{+2}$ and O$^{+3}$ respectively. Therefore the efficiency of the mechanisms  proposed to excite these lines (starlight or Bowen) would be low.

In addition to the results described above, we have found that the {\it FWHM}(ORLs) are in general smaller than the {\it FWHM}(CELs). This is highly significant in particular in objects where two components are detected for the lines, because in this case the red and blue components do not include the effect of expansion, and are affected only by thermal broadening, turbulence and physical structure. The smaller {\it FWHM} of ORLs can be interpreted as ORLs being emitted in zones of lower  temperature than CELs.

Considering the results obtained here for  the majority of our objects and including  the halo PN reported by \citet{otsuka:10} which also shows incompatibility in the kinematics of CELs and ORLs, it seems evident that in most PNe the gas emitting CELs and the gas emitting ORLs have different spatial distributions.

This is consistent with the analysis by \citet{richer:13} for NGC\,7009 who claim that two different plasmas with different conditions and distributions (one emitting CELs and the other, richer and colder, emitting ORLs)  are present in this nebula. The possibility that ORLs and CELs are emitted in different zones of the nebula has been recently reinforced by \citet{richer:17} who analysed the kinematics of the  \ion{C}{ii} $\lambda$6578 recombination line in comparison with the kinematics of H$\alpha$ and the [\ion{N}{ii}] collisionally excited lines in 76 PNe finding that the emission of the \ion{C}{ii} line comes from a volume more internal than what is expected in a plasma in ionisation equilibrium. 

In this work, based on the analysis of the kinematics of CELs and ORLs emitted by many ions,   we conclude that the presence of multiplasma components seems to occur commonly in PNe. Several PN central stars of the objects analysed here, seem to have had at least two different ejections. The first one corresponds to the H-rich plasma  with chemical abundances given by CELs  and whose kinematics behaves accordingly with a structure in ionisation equilibrium and affected by interacting winds as it is predicted by hydrodynamical models,  and a second ejection where the gas is slightly richer in heavy elements than the initial one and that lies inside the previous one.  In several cases this gas, emitting mainly ORLs,  does not behave as a plasma following the hydrodynamic modelling of PNe ionisation structures.

The youngest objects in our sample (those very compact and with a high electron density) show the lowest V$_{exp}$, smaller than 20 km s$^{-1}$. This is specially true for V$_{exp}$ from ORLs.  

Most of PNe studied here have moderate ADFs except NGC\,7009 which shows an  ADF of 5. No correlation seems to exist between the ADF and the difference in kinematics between CELs and ORLs. It would be interesting to analyse the kinematics of CELs and ORLs in  PNe with a very high ADF.  It has been demostrated that at least some PNe with a close binary star show a ver large ADF and extremely rich material forming  knots near the binary central star \citep{corradi:15,garcia:16}. These knots most probably have a kinematics very different from the one of the normal plasma, e.g., see the case of A\,78 by \cite{medina:00}.

The different behaviour in the kinematics of CELs and ORLs does not seem related with the chemical abundances, the excitation of the nebula or the central star (although most of our PNe are ionised by a [WC] star, the ones ionised by {\it wels} or normal stars show the same dissimilar kinematics).

Our results together  with the ones by \citet{richer:13,richer:17} and other authors show the coexistence of different plasmas with different abundances and different physical conditions in PNe. The question relative to what are the real abundances in the plasma of a PNe, if that derived from CELs or that derived from ORLs, seems to have meaning no more, because the different plasmas have their own characteristics.

\section*{Acknowledgements}

This work received finantial support from UNAM DGAPA-PAPIIT IN103117.  J.S.R.-G.  and F.R.-E. acknowledge scholarship from CONACyT M\'exico. This work received partial support from the Spanish Ministry of Economy and Competitiveness (MINECO) under grant AYA2015-65205-P.
J. G.-R.  acknowledges support from Severo Ochoa Excellence Program (SEV-2015-0548) Advanced Postdoctoral Fellowship. 

%%%%%%%%%%%%%%%%%%%%%%%%%%%%%%%%%%%%%%%%%%%%%%%%%%



%\bibliographystyle{mnras}


%%%%%%%%%%%%%%%%%%%%%%%%%%%%%%%%%%%%%%%%%%%%%%%%%%

% Don't change these lines
\bsp	% typesetting comment
\label{lastpage}
\end{document}